\documentclass[twocolumn,floatfix,superscriptaddress]{revtex4}
\usepackage{color}

\usepackage{graphicx}
\usepackage{amsmath}
\usepackage{latexsym}
\usepackage{epstopdf}
\usepackage{amsmath}
\usepackage{amssymb,amsmath}

\newcommand{\beq}{\begin{equation}}
\newcommand{\eeq}{\end{equation}}

\begin{document}

\title{Geometrical exponents of contour loops on synthetic multifractal rough surfaces: multiplicative hierarchical cascade $p$ model}
\author{S. Hosseinabadi,$^1$ M.~A.~Rajabpour,$^2$ M. Sadegh Movahed,$^{3,4,6}$, S. M. Vaez Allaei$^{5,6}$\\
$^1$ Department of Physics, Alzahra University, P.O.Box 19938, Tehran 91167, Iran\\
$^2$ SISSA and INFN, \textit{Sezione di Trieste},  via Bonomea 265, 34136 Trieste, Italy\\
$^3$ Department of Physics, Shahid Beheshti University, G.C., Evin, Tehran 19839, Iran\\
$^4$ School of Astronomy, Institute for Studies in theoretical Physics and Mathematics, P.O.Box 19395-5531, Tehran, Iran\\
$^5$ Department of Physics, University of Tehran, Tehran 14395-547, Iran\\
$^6$ The Abdus Salam International Centre for Theoretical Physics, Strada Costiera 11, I-34013 Trieste, Italy\\}

\vskip 1cm

\begin{abstract}

In this paper, we study many geometrical properties of contour loops to characterize the morphology of 
synthetic multifractal rough surfaces, which are generated by multiplicative hierarchical cascading processes. To 
this end, two different classes of multifractal rough surfaces are numerically simulated. As the Þrst group, singular 
measure multifractal rough surfaces are generated by using the $p$ model. The smoothened multifractal rough 
surface then is simulated by convolving the Þrst group with a so-called Hurst exponent, $H^*$ . The generalized 
multifractal dimension of isoheight lines (contours), $D(q)$, correlation exponent of contours, $x_l$ , cumulative 
distributions of areas, $\xi$, and perimeters, $\eta$, are calculated for both synthetic multifractal rough surfaces. Our 
results show that for both mentioned classes, hyperscaling relations for contour loops are the same as that of 
monofractal systems. In contrast to singular measure multifractal rough surfaces, $H^*$ plays a leading role in 
smoothened multifractal rough surfaces. All computed geometrical exponents for the Þrst class depend not only 
on its Hurst exponent but also on the set of $p$ values. But in spite of multifractal nature of smoothened surfaces 
(second class), the corresponding geometrical exponents are controlled by $H^*$, the same as what happens for 
monofractal rough surfaces. 

\end{abstract}
\maketitle
\section{Introduction}
Random phenomena in nature generate ubiquitously fractal structures
which show self-similar or self-affine properties \cite{sornette2,stanley,ausloos,Feder}.
When the fractal structure of a system is uniform and free of
irregularities, we have monofractal structure.
A monofractal system  can be characterized
by a single scaling law with one scaling exponent in all scales.
For a self-affine surface and interface, this exponent is called
{\it roughness exponent} or {\it Hurst} exponent, $H$. Surface
with larger $H$ seems locally smoother than the surface with smaller $H$ \cite{stanley,ausloos}.

In topics ranging from biology\cite{biology,DFA1}, surface sciences \cite{MultiSurf1,Pradipta,wave,wavelet-based method},
turbulence \cite{smc,benzi84,wt1}, diffusion-limited aggregation \cite{difusion}, bacterial colony
growth \cite{bacteri}, climate indicators \cite{climate} to cosmology \cite{Cosmol},
there are many surfaces and interfaces exhibit multifractal structures.
A multifractal system can be considered as a combination of many
different monofractal subsets \cite{stanley,ausloos}. 
Multifractality manifests itself in systems with different scaling
properties in various regions of the system.
In addition, multifractals can be described by infinite different numbers of
scaling exponents $h(q)$, where $q$ can be a real number. The
appearance of the infinite different numbers ensures that the theoretical
and the numerical study of the multifractal surfaces is more
complicated than those of monofractal ones. Changing one of the $h(q)$'s
can lead to different  feature in the system. One of the important
characteristics of the multifractality is the presence of the
singularity spectrum, $f(\alpha)$, which associates the Husdorff
dimension $f(\alpha)$ to the subset of the support of the measure $\mu$
where the H$\ddot{\rm o}$lder exponent is $\alpha$; 
in other words $f(\alpha)={\rm dim}_{H} \{x|\mu(B_{x}(\epsilon))\sim
\epsilon^{h}\}$, where $B_{x}(\epsilon)$ is an $\epsilon$-box
centered at $x$.

A single scaling exponent can be determined for a
monofractal structure, by use of various methods
\cite{stanley,DFA1,Iraji,other,Chekini,SadeghHermanis,FMTS}. Not only a spectrum of exponents but also different algorithms should be computed for a multifractal feature (power spectral,
distribution method and so on) and these may give different results for a
typical multifractal case \cite{different}. Thus, the better and
more complete theoretical frame work, the better our understanding, providing deeper insight to observational multifractal rough
surfaces.

Recently, isoheight contour lines has been utilized to explore the topography
of  rough surfaces and it exhibited interesting capabilities
\cite{rajab,rohani9,Vaez_Rajabpour,Ghasemi_Rajabpour,Abbas1,Abbas2,KH,KHS}.
The contour plot consists of closed non-intersecting lines in the plane that connects points of equal heights. The fractal properties of the contour loops of
the rough surfaces can be described by just the Hurst exponent \cite{KH,KHS}.
This result was confirmed in different systems with quite different structures
in recent years both experimentally and numerically. Using numerical approach,
the predicted relations were confirmed in glassy interfaces and turbulence  \cite{ZKMM},
in two-Dimensional fractional Brownian motion \cite{Vaez_Rajabpour},
in KPZ surfaces \cite{rohani9} and in discrete scale-invariant rough surfaces
\cite{Ghasemi_Rajabpour}. The predictions were also confirmed by using experimental
data coming from the AFM analysis of WO(3) surfaces \cite{rajab}.

However, although there have been many studies concerning  the
contour lines of monofractal rough surfaces, there ares neither theoretical nor
numerical inferences about the contour lines of multifractal rough surfaces. Because of the presence of
numerous exponents in the multifractal surfaces, theoretical
study of multifractal surfaces seems to be difficult. 
Moreover, in many previous methods,
the exponents determined by fractal analysis, generally provide information about the average
global properties, whereas geometrical analysis addresses information from point to point.
J. Kondev \textit{ et al.} pointed out that geometrical characteristics can discriminate
various monofractal rough surfaces that have a similar power spectrum \cite{KHS,ajay}. Therefore, the geometrical properties may introduce a new opportunity  to
characterize multifractal surface.

It is worth noting that because contour sets are the intersection of a horizontal surface in particular height
fluctuation and do not reflect the full properties of fluctuations in various scales,  it is not trivial  that the geometrical properties
of multifractal rough surfaces based on isoheight nonintersecting feature behave in a  multifractal manner as well.
Therefore we use a new approach to investigate these processes.

In this paper, we try to investigate multifractal structures utilizing contour loops. We study the multifractal
properties of a particular kind of multifractal surfaces. Two different type of synthetic multifractal rough surfaces, namely singular measure and smoothened features,  are generated.
Two mentioned types have a multifractal nature. Despite the complexity nature of the model, hyperscaling relation is satisfied for both categories. 
In addition, from contouring analysis, all of geometrical exponents
for various smoothened multifractal rough surfaces are controlled by
corresponding so-called Hurst exponent,  $H^*$,
This is also is the same as what happens for mono-fractal cases. However,  for a singular measure multifractal rough surface, geometrical exponents depend on the set of $p$values that  is
used to generate the underlying rough surface based on multiplicative cascade model.

The structure of the paper is as follows: In the next section we will review multifractal rough surfaces.
The hierarchical model to generate the surfaces will also given in this section. The multifractal detrended
fluctuation analysis in two dimensions that is used to characterize the multifractal properties of rough surfaces
will be explained in section III.  In  section IV,  nonlinear scaling exponents of multifractal rough surfaces are
introduced.  Section V will be devoted to numerical results for determining  the scaling exponents of the contour loops
in multifractal rough surfaces. In the last section  we will summarize our findings.
\begin{figure}[t]
    \centering
        \includegraphics[width=0.39\textwidth]{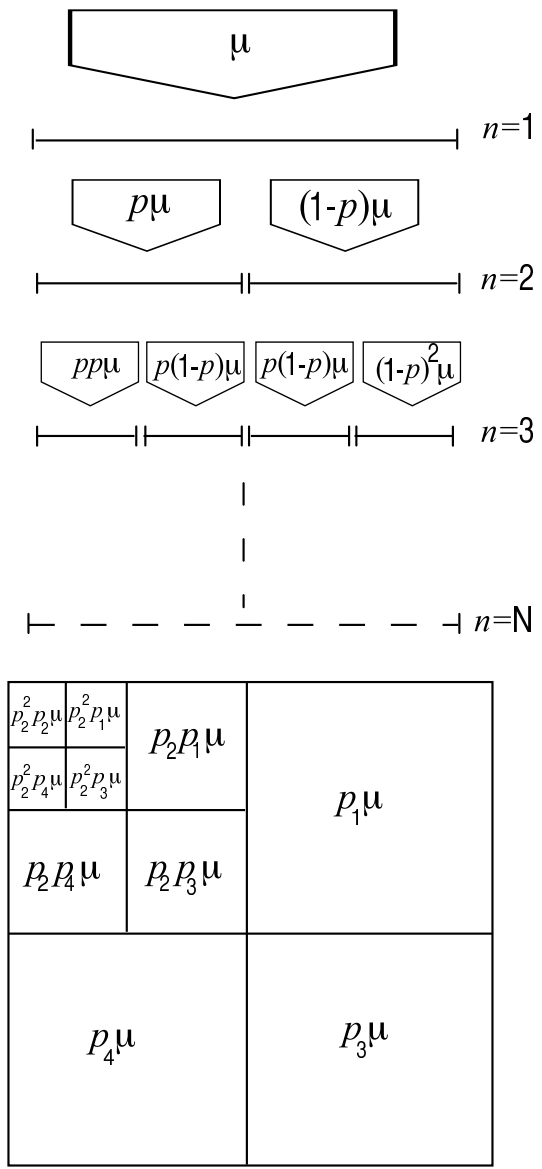}
        \caption{ Upper panel: Different steps of generating  multifractal rough surface in  one Dimension.
        Lower panel:  The  same steps for multifractal rough surface  in  two dimensions [11].} \label{one}
\end{figure}

\section{Multifractal rough surface synthesis}

Recently there has been an increasing interest in the notion of
multifractality because of its extensive applications in different areas
such as complex systems, industrial and natural phenomena.
Dozens of methods for the synthesis of multifractal measures or
multifractal rough surfaces have been  invented. One of the most
common methods thath can be followed deterministically and
stochastically is the multiplicative cascading process
\cite{smc,Feder,Biolog,FMTS}.  Some of these synthesis
methods are known as the random $\beta$ model \cite{benzi84}, $\alpha$ model
\cite{scher884}, log-stable models, log-infinitely divisible cascade
models \cite{scher87,scher97}  and $p$ model \cite{smc}. They
were successfully applied in the studies related to rain in
one dimension, clouds in two dimensions and landscapes in
three dimensions as well as many other fields
\cite{scher87,scher89,scher97,wilson91}.

The  $p$ model  method was proposed to mimic the kinetic energy
dissipation field in fully developed turbulence \cite{smc}.  The
so-called $p$ model represents the spatial version of weighted
curdling feature and is known as conservative cascade . It is based on Richardson's picture of energy
transfer from cores to fine scales base on splitting eddies in a 
random way \cite{frisch} In this model there is no divergency in
corresponding moments in contrast to the so-called hyperbolic of
$\alpha$ model \cite{smc,frisch1}

 On the other hand, many scaling exponents of mentioned model can be determined analytically; therefore, it is a proper method to simulate synthetic multifractal processes ranging from surface sciences and astronomy to high energy physics such as cosmology and particle physics e.g QCD parton shower cascades, and cosmic microwave background radiation \cite{ochs,brax,perivol}. In the context of $p$ model simulating a synthetic   one-dimensional data set,
consider an interval with size $L$. Divide $L$ into two
parts with equal lengths. The value of  the left half corresponds to the fraction $0\leq p\leq 1$ of
a typical measure $\mu$ while the right hand segment is associated to  the remaining
fraction $(1-p)$. By increasing the resolution
to $2^{-n}$, the multiplicative process divides the population in
each part in the same way (see the upper panel of Fig. \ref{one}).

\begin{figure}[t]
    \centering
\includegraphics[width=0.23\textwidth]{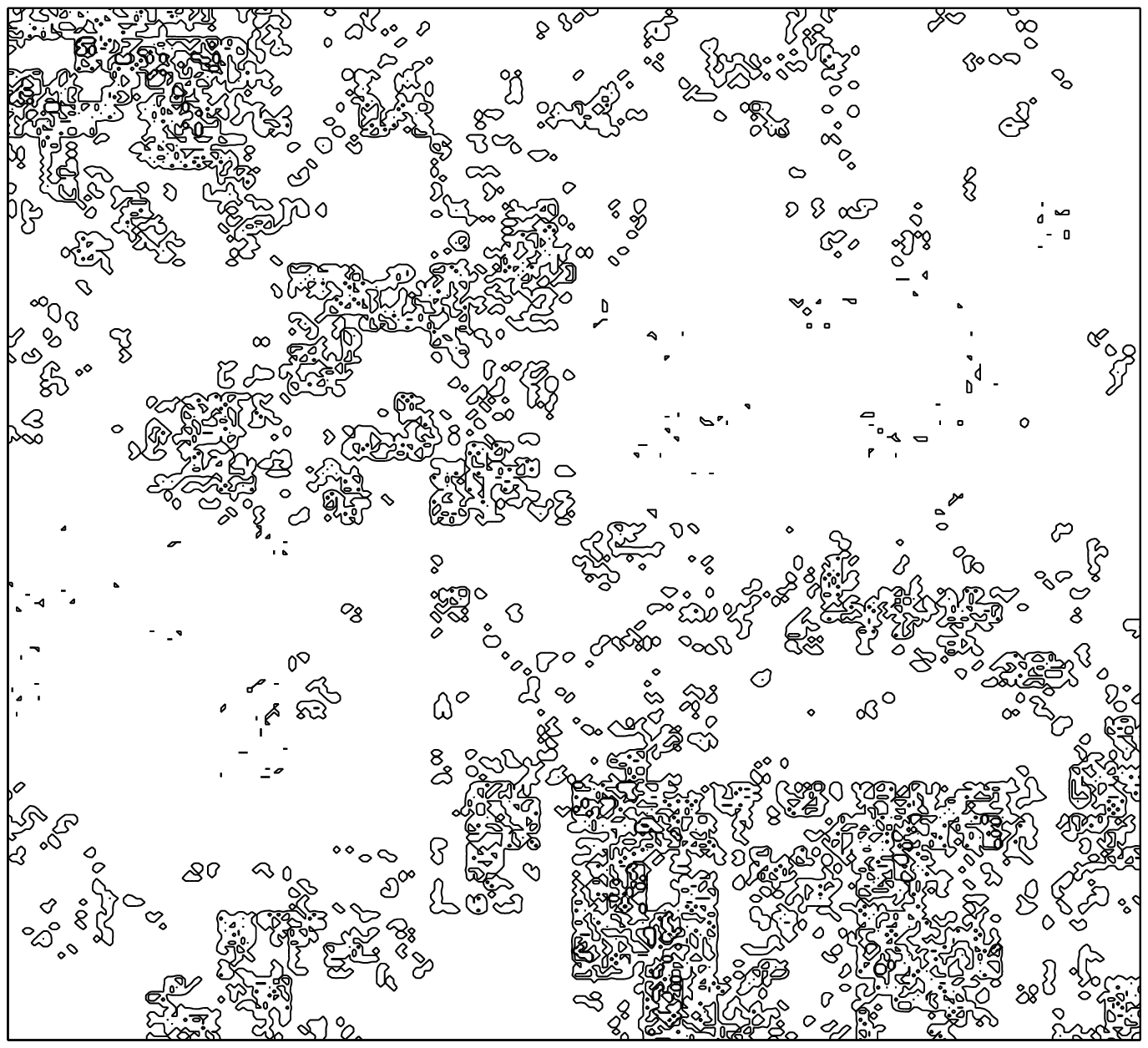}
\includegraphics[width=0.23\textwidth]{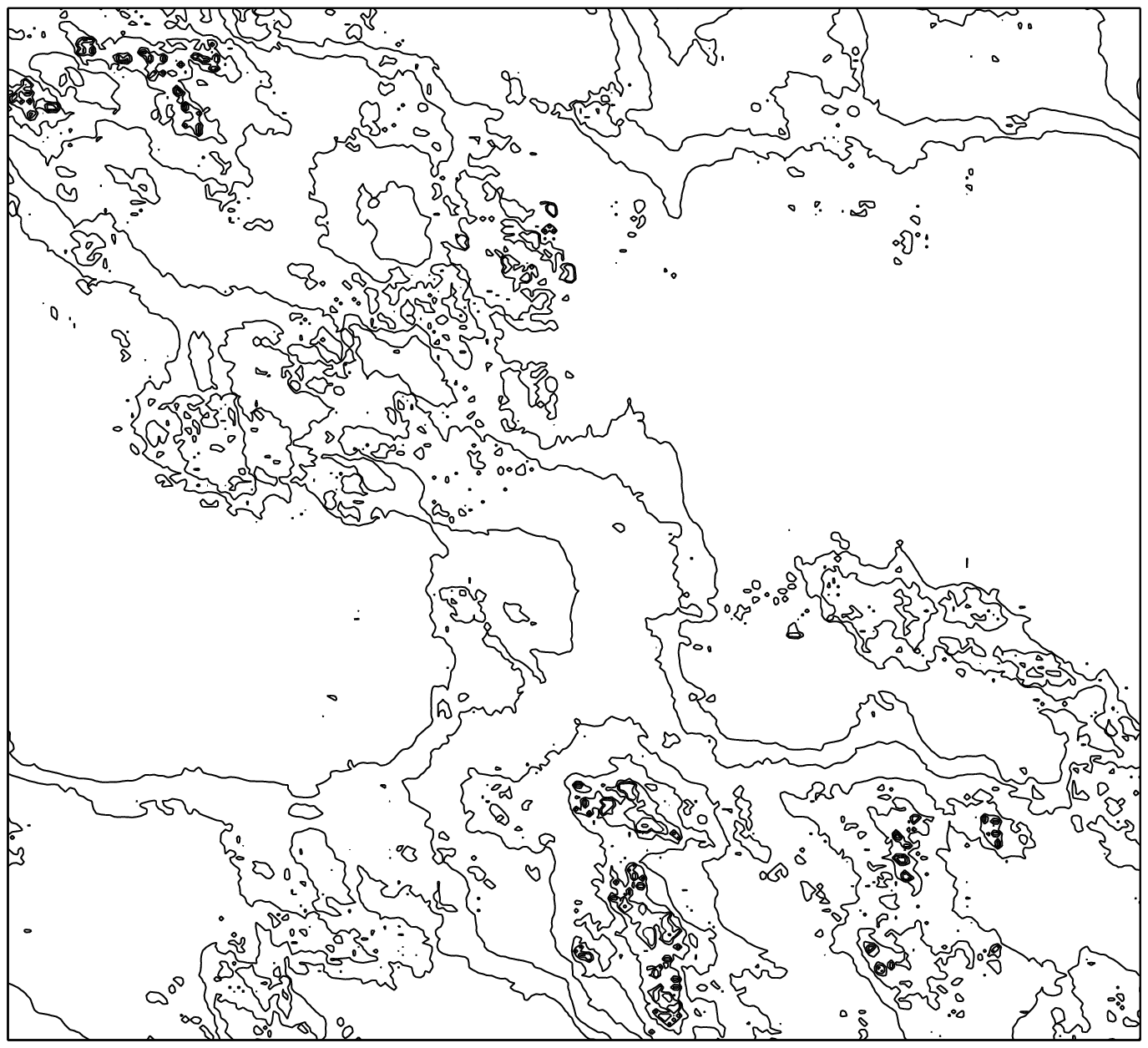}
\caption{Left: Contour plot at some typical levels of a singular  multifractal rough surface generated by binomial
cascade multifractal method with $p=0.22$($H= 0.803$).
The right panel indicates the contour lines of the same surface convolved with $H^*=0.700$. The system size is
$256\times 256$.} \label{Figure:1}
\end{figure}

\begin{figure}[t]
    \centering
    \includegraphics[width=0.5\textwidth]{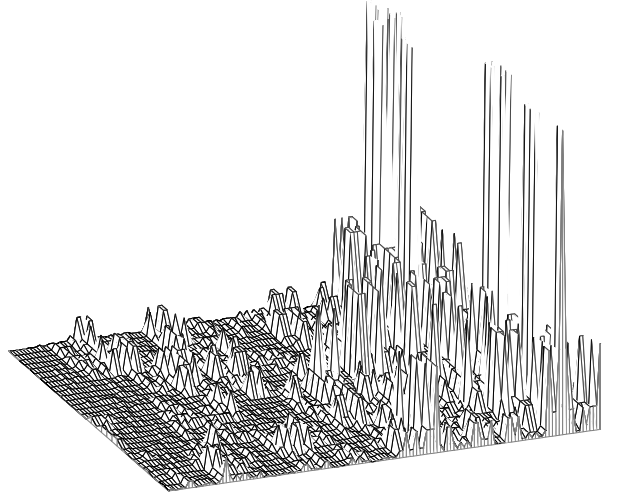}
\includegraphics[width=0.5 \textwidth]{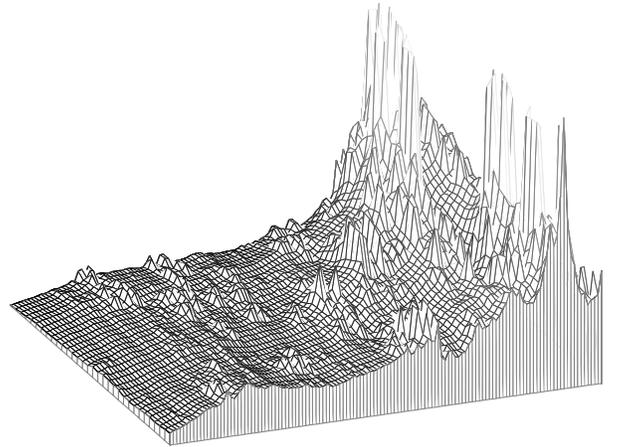}
\caption{Upper panel: A part of height fluctuations of singular
measure mentioned in Fig. 2. Lower panel: The same surface convolved with $H^*=0.700$. }\label{figrough}

\end{figure}

To simulate a mock multifractal rough surface in two dimensions,
one can follow the same procedure as above. Starting from a square,
one breaks it into four sub-squares of the same sizes. The
associated measures for each cell at this step are  $p_1\mu$ for the
upper right cell, $p_2\mu$ for the upper left cell, $p_3\mu$ for the
lower right cell and $p_4\mu$ for the lower left cell. The
conservation of probability at each cascade step is
$p_1+p_2+p_3+p_4=1$. 
This partitioning and redistribution process
repeat and we obtain after many generations, say $n$,  $2^n\times
2^n$ cells of size $l/L=2^{-n}$ (see lower panel of Fig. \ref{one}).
In the stochastic approach, the fraction of measure for each
sub-cell at an arbitrary generation is determined by a random
variable ${\mathcal{A}}$ with a definite probability distribution
function $P({\mathcal{A}})$. By redistribution of measure, based on
independent realization of the random ${\mathcal{A}}$ at smaller
scales, one can generate a random singular measure over a substrate
with size $L\times L$ as

\begin{equation}\label{singular1}
\mu_n( \mathbf{r};l)=\mu\prod_{i=1}^{n(l)}{\mathcal{A}_i( \mathbf{r})}, \quad\quad n(l)=\log_{2}\left(\frac{ L}{l}\right) \rightarrow \infty,
\end{equation}
where $ \mathbf{r}$ shows the coordinate of the underlying cell with size $l$.
In this work, we rely on the stochastic version of
the cascade $p$ model to generate  the synthetic two-dimensional
multifractal rough surface (see Figs. \ref{Figure:1} and \ref{figrough}).  The probability distribution function for our approach is given by
\begin{eqnarray}
P({\mathcal{A}})&=&\frac{1}{4}[\delta({\mathcal{A}}-{\mathcal{A}}_1)+\delta({\mathcal{A}}-{\mathcal{A}}_2)\nonumber\\
&&+\delta({\mathcal{A}}-{\mathcal{A}}_3)+\delta({\mathcal{A}}-{\mathcal{A}}_4)],
\end{eqnarray}
where
\begin{eqnarray}
{\mathcal{A}}_1&=&p_1, \quad {\mathcal{A}}_2=p_2,\nonumber\\
{\mathcal{A}}_3&=&p_3, \quad {\mathcal{A}}_4=p_4.
\end{eqnarray}
The so-called multifractal scaling exponent, $\tau(q)$, and the  generalized Hurst exponent, $h(q)$,
are quantities that  represent the multifractal behaviors of rough surfaces (see section III for more details).
For $p$ model cascade, these exponents can be calculated explicitly.  The scaling exponent $\tau(q)$ is defined via
partition function as
\begin{equation}\label{partition11}
Z_q(l)=\lim_{l\rightarrow 0}\sum_{i=1}^{n(l)}|P({\mathcal{A}}_i,l)|^q\sim l^{\tau(q)}.
\end{equation}
Using the value of $P({\mathcal{A}})$, e.g. for binomial cascade model
$P({\mathcal{A}})=\frac{1}{2}[\delta({\mathcal{A}}-p)+\delta({\mathcal{A}}-(1-p)]$, one finds
\begin{eqnarray}
\tau(q)&=&\lim_{l\rightarrow 0}\frac{\log(Z_q(l))}{\log(l)}\nonumber\\
&=&(E-1)(q-1)-\log_2(p^q+(1-p)^q),
\end{eqnarray}
where $E$ is the dimension of the geometric support, where for our rough surfaces is $E=2$.
For generalized $p$ model, the analytic expression of  multifractal scaling exponent in two dimensions is given by \cite{Kadanoff}
\begin{equation}\label{tauqp}
\tau(q)=-\log_2(p_1^q+p_2^q+p_3^q+p_4^q).
\end{equation}
One can use the above theoretical expression to get the  most relevant quantities of the multifractal behavior and
check the reliability and the robustness of numerical method.

Recently, factorial moments, G-moments, correlation integrals, void
probabilities, combinants and wavelet correlations have been used to
examine many interesting feature of multiplicative cascade processes
\cite{casc}. But there is some ambiguity in properties of such
processes that  represent multifractal phenomena. On the other
hand, sensitivity and accuracy of results are method dependent;
consequently,  it is highly proposed to simultaneously use various
tools in order to ensure the reliability of given results for
underlying multifractal rough features. Moreover, to make a relation
between experimental data and simulation, generally, we require more
than one characterization \cite{KHS,gomez}

In the next section, to investigate the multifractal properties of simulated rough surfaces in two dimensions,
we will introduce the so-called multifractal detrended fluctuation analysis.


\section{Multifractality of synthesis rough surface}

There are many different
methods to determine the multiscaling properties of real as well as synthetic multifractal
surfaces such as spectral analysis \cite{SA1},
fluctuation analysis \cite{FA}, detrended fluctuation analysis
(DFA) \cite{DFA1,DFA2,DFA3}, wavelet transform module maxima (WTMM)
\cite{wave,wavelet-based method,wt1,wt2,wt3} and discrete wavelets \cite{kantelh95,kantelh96}.
For  real data sets and in the presence of noise,
the multifractal DFA (MF-DFA) algorithm gives very reliable results \cite{Biolog,DFA2}.  Since it
does not require the modulus maxima procedure therefore this method is simpler than WTMM, however, it involves
a bit more effort in programming.

In this work,
we rely on the two dimensional multifractal detrended fluctuation analysis (MF-DFA) to determine the spectrum of the
generalized Hurst exponent, $h(q)$. We then compare given results with theoretical prediction to check the reliability of our simulation. Suppose that for a rough surface in two dimensions height of the fluctuations
is represented by
${\mathcal{H}}( \mathbf{r})$ at coordinate $ \mathbf{r}=(i,j)$ with resolution $\Delta$.
The MF-DFA in two dimensions has the following steps \cite{Biolog}

\textit{Step1}: Consider a two dimensional array ${\mathcal{H}}(i,j)$ where $i=1,2,...,
M $ and $j=1,2,..., N $. Divide the ${\mathcal{H}}(i,j)$ into $ M_s\times
N_s$ non-overlapping square segments of equal sizes $s\times s$,
where $M_s=[\frac{M}{s}]$ and $N_s=[\frac{N}{s}]$. Each square segment can be denoted by ${\mathcal{H}}_{\nu,w}$
such that
${\mathcal{H}}_{\nu,w}(i,j)={\mathcal{H}}(l_1+i,l_2+j)$ for $1 \leq i,j \leq s$, where
$l_1=(\nu-1)s $ and $l_2=(w-1)s $.

\textit{Step 2}:  For each non-overlapping segment, the cumulative sum is
calculated by:
\begin{eqnarray}\label{cumulativesum}
Y_{\nu,w}(i,j)=\sum_{k_1=1}^i \sum_{k_2=1}^j
 {\mathcal{H}}_{\nu,w}(k_1,k_2);
\end{eqnarray}
where $1\leq i,j\leq s$.

\textit{Step 3}: Calculating  the local trend for each segments by a least-squares of the profile, linear,
 quadratic or higher order polynomials can be used in the fitting procedure as follows:
\begin{eqnarray}\label{fitting}
{\mathcal{B}}_{\nu,w}(i,j)&=&ai+bj+c,\\
{\mathcal{B}}_{\nu,w}(i,j)&=&ai^2+bj^2+c.
\end{eqnarray}
 Then determine the variance for each segment as follows:
\begin{eqnarray}\label{variance}
{\mathcal{D}}_{\nu,w}(i,j)&=&Y_{\nu,w}(i,j)-{\mathcal{B}}_{\nu,w}(i,j),\\
F^2_{\nu,w}(s)&=& \frac{1}{s^2}\sum_{i=1}^s \sum_{j=1}^s
 {\mathcal{D}}^2_{\nu,w}(i,j).
\end{eqnarray}
A comparison of the results for
different orders of DFA allows one to estimate the type of the
polynomial trends in the surface data.

\textit{Step 4}: Averaging over all segments to obtain the $q$'th order
fluctuation function
\begin{eqnarray}\label{fluctuation function}
 F_{q}(s)= \Big{(}\frac{1}{M_{s}\times N_{s}}
\sum_{\nu=1}^{M_{s}}\sum_{w=1}^{N_{s}}\left[F^{2}_{\nu,w}(s)\right]^{q/2}\Big{)}^{1/q},
\end{eqnarray}
where $F_q(s)$ depends on scale $s$ for different values of $q$. It is easy to see
that $F_q(s)$  increases with increasing $s$. Notice that $F_q(s)$
depends on the order $q$. In principle, $q$ can take any real value except zero. For $q=0$ Eq. (\ref{fluctuation function}) becomes
\begin{eqnarray}\label{fluctuation function0}
 F_{0}(s)= \exp\Big{(}\frac{1}{2M_{s}\times N_{s}}
\sum_{\nu=1}^{M_{s}}\sum_{w=1}^{N_{s}}\ln F^{2}_{\nu,w}(s)\Big{)}.
\end{eqnarray}

For $q=2$ the standard DFA in two dimensions will be retrieved.

\textit{Step 5}: Finally, investigate the scaling behavior of the fluctuation
functions by analyzing log-log plots of $F_q(s)$ versus $s$ for each
value of $q$,
\begin{eqnarray}\label{fluctuation function1}
 F(s)\sim s^{h(q)}\label{f(s)}.
\end{eqnarray}
The  Hurst exponent is given by 
\begin{equation}\label{hurst11}
 H\equiv h(q=2)-1.
 \end{equation}
 Using standard multifractal formalism \cite{DFA2} we have
\begin{equation}\label{tauqq}
\tau(q)=qh(q)-E.
\end{equation}

 It has been shown that for very large scales, $
N/4<s$, $F_q(s)$ becomes statistically unreliable because the number
of segments $N_s$ for the averaging procedure in step 4 becomes
very small \cite{Biolog}. Thus, scales $ N/4<s$ should be excluded from the
fitting procedure of determining $h(q)$. On the other hand one should be careful also about systematic
deviations from the scaling behavior in Eq. (\ref{fluctuation function}) that can occur for the small scales $s<10$.

The singularity spectrum, $f(\alpha)$, of a multifractal rough surface is given by the Legendre transformation of
$\tau(q)$ as
\begin{equation}\label{falpha}
f(\alpha)=q\alpha-\tau(q),
\end{equation}
where $\alpha=\frac{\partial \tau(q)}{\partial q}$. It is well-known that for a multifractal surface, various parts of
the feature are characterized by different values of $\alpha$, causing a set of H$\ddot{\rm o}$lder exponents
instead of a single $\alpha$. The interval of H$\ddot{\rm o}$lder spectrum, $\alpha\in [\alpha_{\rm min},\alpha_{\rm max}]$,
can be determined by \cite{muzy94,muzy95}
\begin{eqnarray}
\alpha_{\rm min}&=&\lim_{q \rightarrow +\infty} \frac{\partial \tau(q)}{\partial q},\\
\alpha_{\rm max}&=&\lim_{q \rightarrow -\infty} \frac{\partial \tau(q)}{\partial q}.
\end{eqnarray}
\begin{figure}[t]
    \centering
     \includegraphics[width=0.45\textwidth]{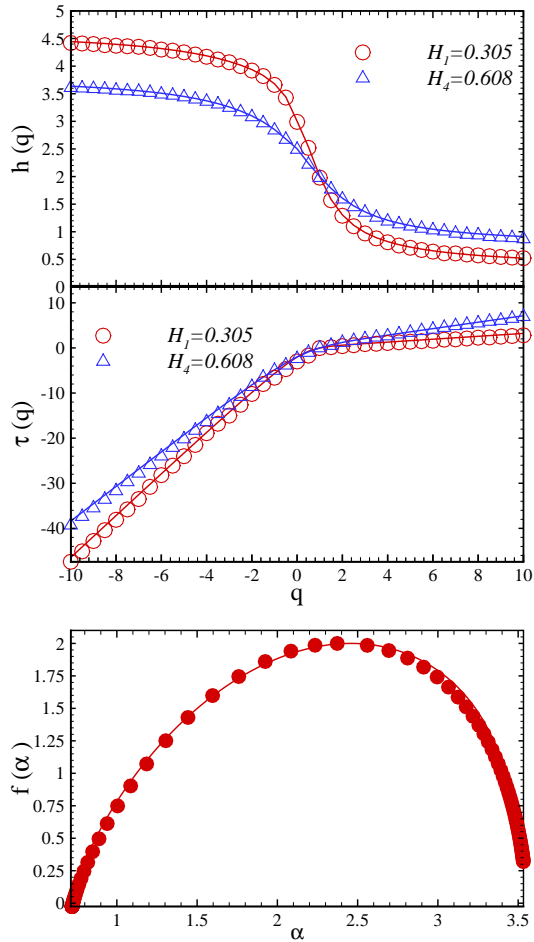}
\caption{(Color online) Diagrams of $h(q)$ (upper panel) and $\tau(q)$ (middle panel) for different surfaces.
We have distinguished different surfaces with their $H_i=h_i(q=2)-1$
coming from the Table I. The subindex ($i \in[1,12]$) of each $H_i$ (Hurst exponent) throughout this paper corresponds to a given set of $p$ values reported in Table I.  The lower panel corresponds to the singularity spectrum of a typical multifractal rough surface
with $H_4=0.608$.
In all diagrams, symbols and solid lines correspond to results given by numerical calculation and  theoretical formula,
respectively. } \label{Figure:2}
\end{figure}

To evaluate the statistical errors due to numerical calculations
we introduce posterior probability distribution function in terms of likelihood analysis.
To this end, suppose the measurements and model parameters to be
assigned by  $\{X\}$ and $\{\Theta\}$, respectively.The conditional probability of the model
parameters for a given observation is as follows (posterior)
\begin{equation}
P(\Theta|X)=\frac{{\mathcal{L}}(X|\Theta)P(\Theta)}{\int
{\mathcal{L}}(X|\Theta)P(\Theta)d\Theta}.
\end{equation}
here ${\mathcal{L}}(X|\Theta)$ and $P(\Theta)$ are called
Likelihood and  prior distribution, respectively. The prior distribution containing all initial constraints
regarding model parameters. 
Based on 
the central limit theorem, Likelihood function can be given by a product
of gaussian functions as follows:
\begin{equation}\label{likelihood1}
\ln{\mathcal{L}}(X|\Theta)\sim \frac{-\chi^2(\Theta)}{2},
\end{equation}
where e.g., for determining $h(q)$ we have  $\{X\}:\{F_q(s)\}$ as observations and $ \{\Theta\}:\{h(q)\}$ as
free parameter to be determined. Also
\begin{equation}
\chi^2(h(q))=\sum_s \frac{[F_{{\rm obs.}}(s)-F_{{\rm the.}}(s;h(q))]^2}{\sigma_{{\rm obs.}}^2(s)},
\end{equation}\label{chii}
where  $F_{{\rm obs.}}(s)$ is computed by Eqs. (\ref{fluctuation function}) and (\ref{fluctuation function0}).  $F_{{\rm the.}}(s;h(q))$ is
the fluctuation functions given by Eq. (\ref{f(s)}). The observational error is $\sigma_{{\rm obs.}}(s)$.
By using the fisher matrix,  one can evaluate the value of the error-bar at
$1\sigma$ confidence interval of $h(q)$ \cite{fisher}
\begin{equation}
\mathcal{F}(q)\equiv \left \langle \frac{\partial^2\ln \mathcal{L} }{\partial h(q)^2}\right \rangle
\end{equation}
and
\begin{equation}
\sigma(q)\simeq\frac{1}{\sqrt{\mathcal{F}(q)}}
\end{equation}

Finally we report the best value of the scaling exponent at $1\sigma$
confidence interval according to $h(q)\pm \sigma (q)$. 
Using the method mentioned in the previous section, we simulated
multifractal rough surfaces and  checked their multifractality
nature by using the spectrum of $h(q)$. Figure
\ref{Figure:2} shows the generalized Hurst exponent and $\tau(q)$
as a function of $q$ 
for various values of measure sets reported
in Table \ref{Tab1}. The subindex ($i \in[1,12]$) of each $H_i$ (Hurst exponent) throughout this paper corresponds to a given set of $p$ values reported in Table \ref{Tab1}. In addition the singularity spectrum of a
typical simulated multifractal rough surface has been shown in the
lower panel of Fig. \ref{Figure:2}. The $q$ dependence of $h(q)$
as well as the extended range of singularity spectrum demonstrate
the multifractality nature of synthesis rough surfaces.
Theoretical predictions of $\tau(q)$, $h(q)$ and $f(\alpha)$ shown
by the solid lines in the corresponding plots, are given by Eqs.
(\ref{tauqp}), (\ref{tauqq}) and (\ref{falpha}), respectively.
There is a good consistency between theoretical predictions and
computational values.

\begin{figure}[t]
    \centering
        \includegraphics[width=0.50\textwidth]{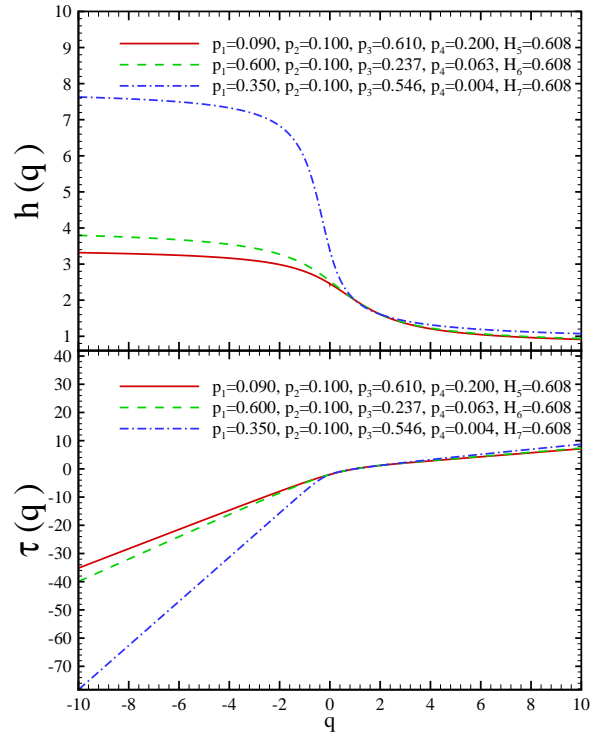}
\caption{(Color online)  The multifractal spectrum of surfaces produced by different sets of $p$ values but with the same $h(q=2)$ up to our numerical precision.} \label{pp}
\end{figure}

Before going further it is worth mentioning that in the cascade $p$ model  for various sets of $p$ values which have
the same $h(q=2)$, in principle, there exist different $h(q)$ spectrums. To show this point,
we fixed the value of $\tau(q=2)$ in Eq. (\ref{tauqp}) and by having e.g., $p_1$ and $p_2$,
one can compute the rest of $p$ values according to normalization of $p$'s. In  Fig. \ref{pp}
we show MF-DFA results of various sets of $p$ values causing the same so-called $h(q=2)$ exponents.
Subsequently, it is expected that for characterizing the geometrical properties of underlying surfaces,
one must take into account full spectrum of generalized Hurst exponents.

\begin{table}[t]

\begin{center}
\begin{tabular}{|c|c|c|c|c|c|}\hline
 {\rm Hurst exponent}   & $p_{1}$ & $p_{2}$ & $p_{3}$& $p_{4}$ \\
\hline
$H_1=0.305$ & $0.040$ & $0.800$ &$0.080$& $0.080$\\
\hline
$H_2=0.404$ & $0.100$ & $0.740$ &$0.080$& $0.080$\\
\hline
$H_3=0.504$ & $0.120$ & $0.680$ &$0.110$& $0.090$\\
\hline
$H_4=0.608$ & $0.190$ & $0.610$ &$0.130$& $0.070$\\
\hline
$H_5=0.608$ & $0.090$ & $0.100$ &$0.610$& $0.200$\\
\hline
$H_6=0.608$ & $0.600$ & $0.100$ &$0.237$& $0.063$\\
\hline
$H_7=0.608$ & $0.350$ & $0.100$ &$0.546$& $0.004$\\
\hline
$H_8=0.706$ & $0.210$ & $0.550$ &$0.130$& $0.110$\\
\hline
$H_9=0.802$ & $0.220$ & $0.480$ &$0.200$& $0.100$\\
\hline
$H_{10}=0.697$ & $0.120$ & $0.180$ &$0.560$& $0.140$\\
\hline
$H_{11}=0.806$ & $0.160$ & $0.180$ &$0.170$& $0.490$\\
\hline
$H_{12}=0.906$ & $0.410$ & $0.200$ &$0.210$& $0.180$\\
\hline
    \end{tabular}
\end{center}
\caption{\label{Tab1}The $p$ values used for construction of surfaces
with various Hurst exponents, $H_i=h_i(q=2)-1$. The subindex ($i \in [1,12]$) of $H_i$ represents the label of different sets of $p$ values.}
\end{table}

\begin{figure}[t]
    \centering
        \includegraphics[width=0.50\textwidth]{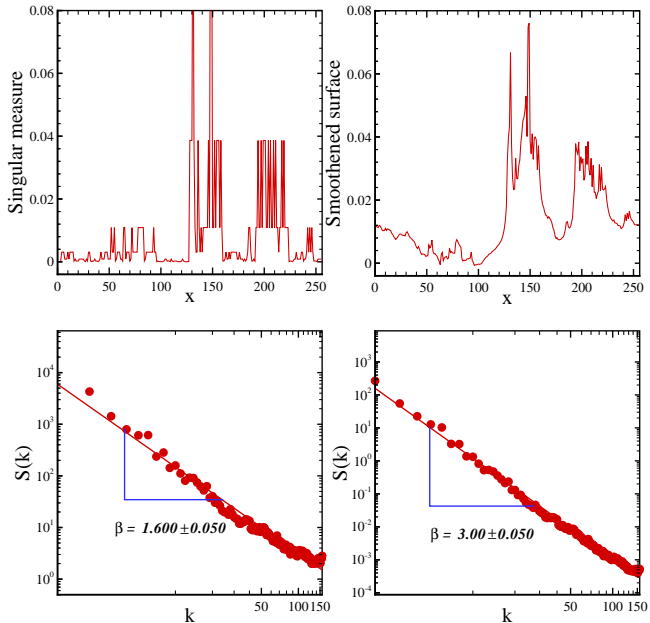}
\caption{(Color online)  Upper panel:  Pprofile of singular (left)
and smoothened (right) multifractal rough surfaces along a typical
horizontal cut in Fig. 2. Lower panel:   Spectral density of
mentioned mock rough surfaces. The solid lines in the lower panel
corresponds to a power law fitting function and symbols are given by
numerical calculation. Here we took $H^*=0.700$.} \label{power}
\end{figure}

 It must be pointed out that the generated surfaces have some discontinuities (see Fig. \ref{Figure:1}).
 To make them smooth, a proper way is using  fractionally integrated singular cascade
(FISC) method \cite{wavelet-based method}. In this method, the multifractal measure is
transformed into a smoother multifractal rough surface by filtering
the singular multifractal measure [$\mu(\mathbf{r})$, (Eq. (\ref{singular1}))] in the Fourier space as
\begin{eqnarray}\label{smooth function}
 {\mathcal{H}}( \mathbf{r})=\mu( \mathbf{r})\otimes| \mathbf{r}|^{-(1-H^*)},
\end{eqnarray}
 where $\otimes$ is the convolution operator
and $H^*\in (0,1)$ is the order of smoothness (see the right panel of Fig. \ref{Figure:1} and lower panel of
Fig. \ref{figrough}). In this case
$\tau_f(q)$ reads as
\begin{eqnarray}\label{tau}
 \tau_f(q)=\tau(q)+qH^*,
\end{eqnarray}
where $\tau(q)$ is given by Eq.(\ref{tauqp}).  Using the correlation
function, $C(|\mathbf{r}|)\sim|\mathbf{r}|^{-\gamma}$, and its
Fourier transform one can derive the power spectrum scaling exponent
$\beta$ of the singular as well as the smoothened synthetic multifractal
surfaces.  To this end we demand the scaling behavior for power
spectrum to be
\begin{equation}
S(k)\sim |\mathbf{k}|^{-\beta},
\end{equation}
where $\mathbf{k}=(k_x,k_y)$, $k_x=\frac{2\pi}{\Delta\times N}i$, $k_y=\frac{2\pi}{\Delta \times N}j$ and $(i,j)$ run from $1$ to
$N=L/\Delta$ (the pixel of system size).
Subsequently the power spectrum scaling exponent is given by \cite{wavelet-based method}
\begin{eqnarray}
\beta&=&1+2H^*-\log_2(p^2+(1-p)^2).
\end{eqnarray}
To make more sense, in Table
\ref{exponents} we collected the correlation and power spectrum exponents  of
stochastic processes in one and two dimensions .
\begin{table}[htp]
\begin{center}
\begin{tabular}{|c|c|c|c|c|c|}\hline
   Exponent & 1D-fGn & 1D-fBm &2D-Cascade& 2D-fBm \\
\hline
$\gamma$ & $2-2H$ & $-2H$ &$1-2H$& $-1-2H$\\
\hline
$\beta$ & $2H-1$ & $2H+1$ &$2H$& $2H+2$\\
\hline
    \end{tabular}
\end{center}
\caption{\label{exponents}The most relevant exponents concerning stochastic processes in one and two dimensions.}
\end{table}

\begin{figure}[t]
    \centering
        \includegraphics[width=0.50\textwidth]{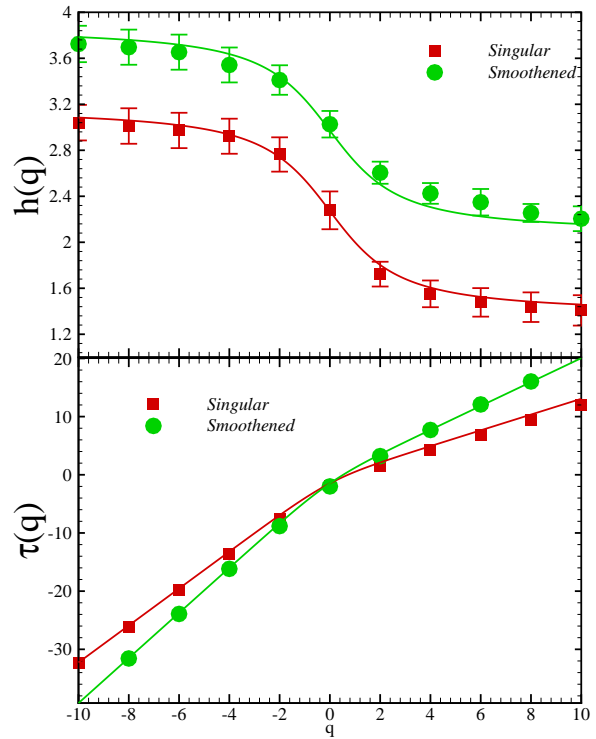}
\caption{(Color online)  Generalized Hurst exponent of singular measure for $H_9=0.802$ (square symbols) and that of convolved with $H^*=0.700$ (circle symbols). The solid lines are from the theory.} \label{hqss}
\end{figure}

Figure \ref{power} indicates one dimension profiles obtained along a
typical horizontal cut in Fig. \ref{Figure:1} for singular and
smoothened multifractal rough surfaces. The lower panels of Fig.
\ref{power} show the power spectrums of simulated rough surfaces.
The convolution does not change the multifractality nature of
singular measure (see  Fig. \ref{hqss}). In this plot one can see
that the synthetic smoothened surface remains multifractal.

\section{Geometrical exponents of contour loops}

For a given  multifractal rough surface with the height
${\mathcal{H}}(\mathbf{x})$, a level set
${\mathcal{H}}(\mathbf{x})={\mathcal{H}}_0$ for different values
of ${\mathcal{H}}_0$ consists of many closed non-intersecting
loops. These loops are recognized as contour loops.  The contour
loop ensemble corresponds to contour loops of various level sets.
In Fig. \ref{Figure:1} we plotted a set of contour loop at some
typical levels for singular multifractal rough surface and
corresponding convolved surface with $H^*=0.700$. The loop length
$s$ can be defined as the total number of unit cells constructing
a contour loop multiplied by lattice constant $\Delta$.  The
radius of a typical loop is represented by $R$ and it is the side
of the smallest box that completely enwraps the loop.  For a mono-fractal surface, these
loops are usually fractal and their size distribution is
characterized by a few scaling functions and scaling exponents.
For example the contour line properties can be described by the
loop correlation function $G(\mathbf{r})$.  The loop correlation
function measures the probability that the two points  separated
by the distance $\mathbf{r}$ in the plane lie on the same contour.
Rotational invariance of the contour lines forces  $G(\mathbf{r})$
to depend only on $r=\vert \mathbf{r} \vert$. This function for
the contours on the lattice with grid size $\Delta$ and in the
limit $r\gg \Delta$ has the scaling behavior
\begin{eqnarray}\label{G_scaling}
G(r) \sim r^{-2x_l},
\end{eqnarray}
where $x_l$ is the loop correlation exponent. It was shown numerically \cite{KH, KHS, Ghasemi_Rajabpour} that for all
the known mono-fractal rough surfaces this exponent is superuniversal and equal to $\frac{1}{2}$. A key consequence of
this result is that, the contour loops with perimeter $s$ and radius $R$ of such surfaces are self-similar.
When these lines are scale invariant, one can determine the fractal dimension as the exponent in the perimeter-radius
relation. The relation between contour length $s$ and its radius of gyration $R$ is

\begin{eqnarray}\label{scaling_of_s}
\langle s\rangle (R) \sim R^{D_f},
\end{eqnarray}
where $D_f$ is the fractal dimension and $R$ is defined by
$R^2=\frac{1}{N} \sum_{i=1}^N \Big{[}(x_i-x_c)^2+(y_i-y_c)^2\Big{]}$, with $x_c=\frac{1}{N}\sum_{i=1}^N x_i$
and $y_c=\frac{1}{N}\sum_{i=1}^N y_i$ being the central mass coordinates. The $D_{f}$ is the fractal dimension
of one contour and for mono fractal rough surfaces is given by $D_{f}=\frac{3-H}{2}$ \cite{KHS}.
Depending on what feature of the multifractal rough surface is under investigation  one can get various types of
fractal dimensions. In this paper we introduce the fractal dimension of a isoheight line, $D_f$, and the fractal
dimension of all the level set, $d$.
The generalized form of fractal dimension can be expressed by means of partition function of underlaying feature,
which is contours in this context, as
\begin{eqnarray}\label{generalized fractal dimension}
D(q)=\lim_{l\rightarrow
0}\frac{1}{q-1}\frac{\log(Z_q(l))}{\log(l)},
\end{eqnarray}
where $l$ is the size of the cells that one uses to cover the domain and its minimum value is equal to grid size,
$\Delta$. $Z_q(l)$ is the partition function defined in Eq. (\ref{partition11}) but here it should be constructed by
using contour loops instead of height function and $q$ can be any real number. It is easy to show that $D(q=0)=D_f$ and
$D(q=1)$ corresponds to the so-called entropy of underlying system \cite{Kadanoff}.

For a given self-similar loop ensemble, one can define the probability distribution of contour lengths $\tilde{P}(s)$.
This function is a measure for the total loops with length $s$ and follows the power law
\begin{eqnarray}\label{scaling_of_P}
\tilde{P}(s) \sim s^{-\eta},
\end{eqnarray}
where $\eta$ is a scaling exponent. Another interesting quantity with the scaling property is the
cumulative distribution of the number of contours with area
greater than $A$ which has the following form
\begin{eqnarray}\label{cumulative}
P_{>}(A)\sim A^{-\frac{\xi}{2}}.
\end{eqnarray}
For mono fractal rough surfaces we have $\xi=2-H$. Using the scaling property of the mono-fractal surfaces
it was shown that the three exponents $D_{f}$, $\eta$, $\xi$ and $x_l$ satisfy the following hyperscaling
relations \cite{KH}
\begin{eqnarray}\label{hyperscaling}
D_{f}=\frac{\xi}{(\eta-1)},\\
D_f=\frac{2x_l-2}{\eta-3}.
\end{eqnarray}
Using the above relations  it is easy to get the relation between $\eta$ and Hurst exponent $H$.
Before closing this section, we summarize all of the exponents introduced in this section in Table \ref{tabintroduce}.

\begin{table}[htp]
\begin{center}
\begin{tabular}{|c|c|c|}\hline
Exponent &Relation &Description\\
\hline
$x_l$ & $G(r) \sim r^{-2x_l}$& Loop correlation exponent\\
\hline
$D_f$ & $\langle s\rangle (R) \sim R^{D_f}$& Fractal dimension of a contour loop\\
\hline
$D(q)$ & Eq. (31) & Multifractal dimension\\
\hline
$d$ & $N(l)\sim l^{-d}$& Fractal dimension of all contour set \\
\hline

$\eta$ & $\tilde{P}(s) \sim s^{-\eta}$& Length distribution exponent  \\
\hline
$\xi$ & $P_{>}(A)\sim A^{-\frac{\xi}{2}}$& Area cumulative exponent\\
\hline

    \end{tabular}
\end{center}
\caption{\label{tabintroduce}The relevant exponents introduced in this paper to characterize synthetic multifractal
rough surfaces.}
\end{table}
In the next section we will calculate all mentioned exponents by
using different numerical methods  for singular as well as smoothened
multifractal rough surfaces and we will examine the validity of the
hyperscaling relations in this context.

\begin{figure}[t]
    \centering
        \includegraphics[width=0.50\textwidth]{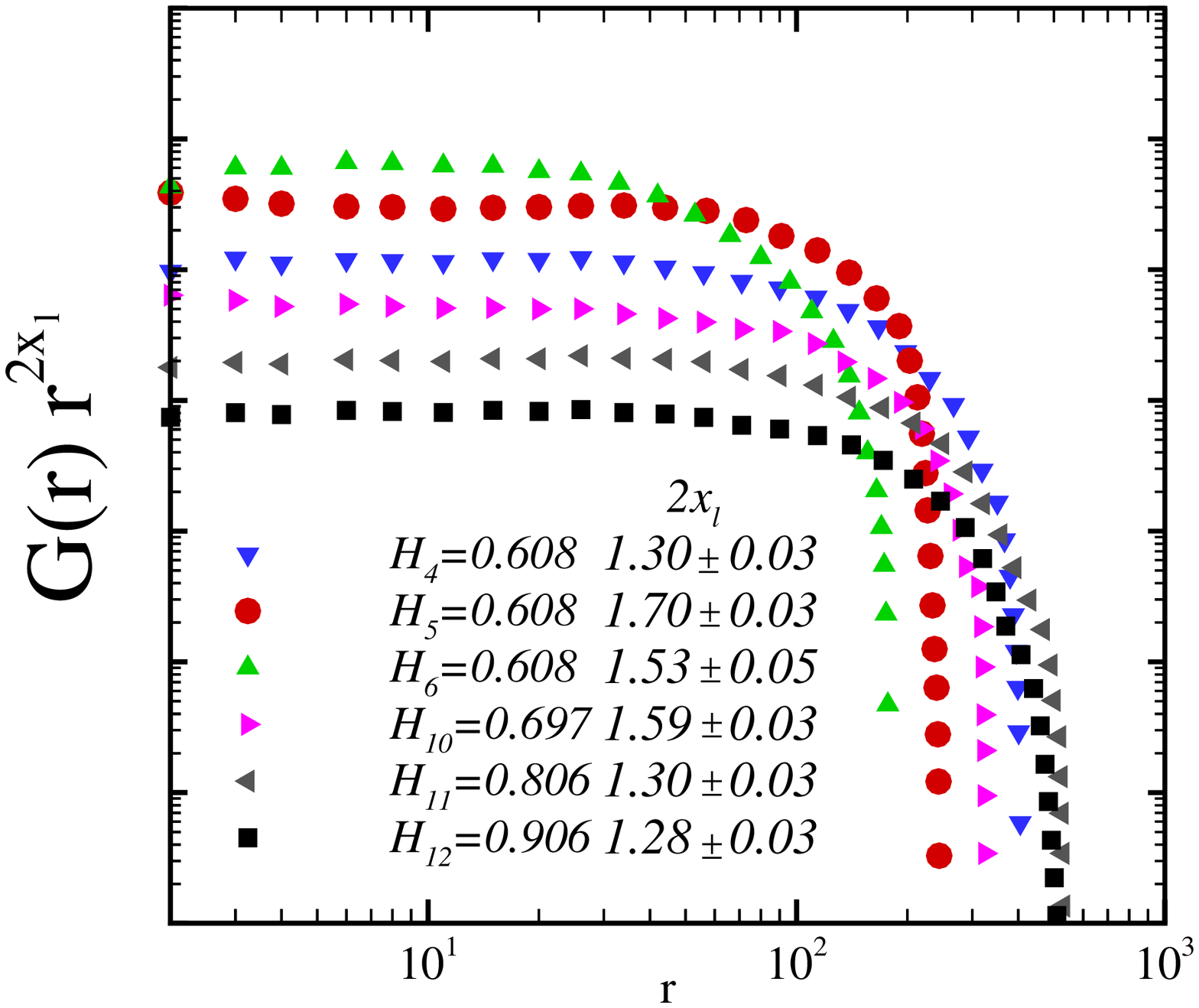}
\includegraphics[width=0.50\textwidth]{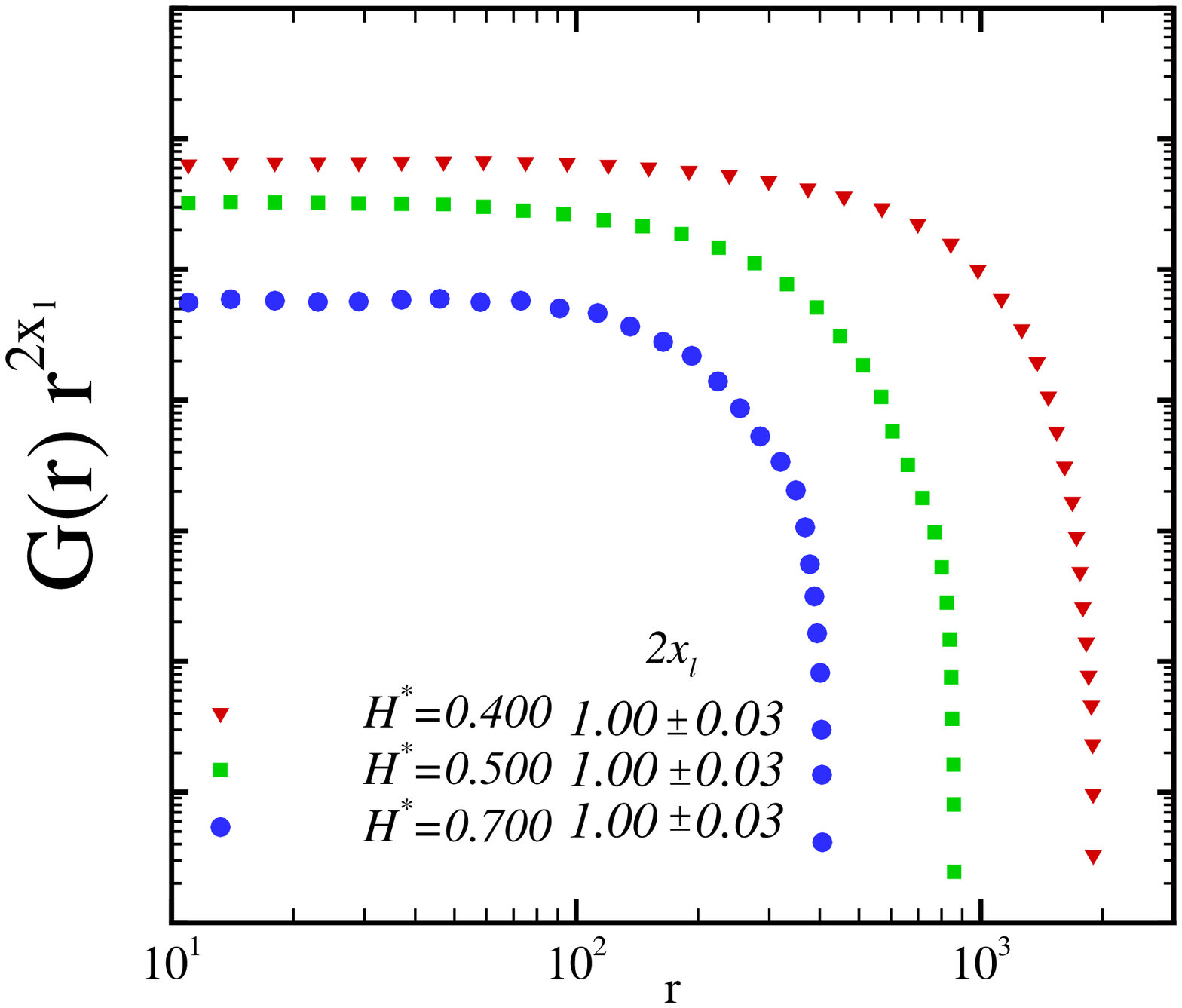}
\caption{(Color online) Log-log diagram of $r^{2x_{l}}G(r)$ versus $r$ for
different Hurst exponents. Upper panel corresponds to singular measure with the sets of $p$ values reported in Table I.
Lower panel is indicates loop correlation function for smoothened multifractal surface for various $H^*$'s. In these figures we shifted the $y$ axis vertically.The system size is $4096\times4096$.}
\label{Figure:5}
\end{figure}

\section{Numerical results}

In order to examine  the geometrical exponents of the contour loops
mentioned in Table \ref{tabintroduce} of synthetic multifractal
rough surfaces,  we have generated multifractal rough surfaces with
different $h(q=2)$'s using the typical measures reported in Table
\ref{Tab1}. We have generated $100$ ensembles of each surfaces with
various sizes ranging from $(2048\times 2048)$ to $(4096\times
4096)$. To extract the contour loops of the mock multifractal rough
surfaces at mean height, ${\mathcal{H}}_0$,   we use two
different methods, the contouring algorithm and Hoshen-Kopelman
algorithm \cite{Vaez_Rajabpour}. According to our results, these two
methods give almost the same results for geometrical exponents. In the next subsections we present
 our numerical results concerning the exponents introduced in the preceding sections.

\subsection{Loop correlation function exponent}

The loop correlation function exponent $x_{l}$ is the most central
exponent in mono-fractal rough surfaces. It is independent of $H$ and
is equal to $\frac{1}{2}$. This result has also been proven
for $H=0$ according to the exact solvable statistical mechanics model for
contours  equivalent to the critical $O(2)$ loop model on the
 honeycomb lattice \cite{nien87,KHS}.

To find the
correlation function from a given loop ensemble for multifractal
rough surfaces, we followed the algorithm described in Ref.  \cite{KHS}.
We calculated the loop correlation function $G(r)$ for our
multifractal rough surfaces (with system size $ 2048 \times 2048$
and averaging is done over $10$ realizations). The log-log diagram
of $G(r)r^{2x_l}$ versus $r$ for different sets of $p$ values ($H_i=h_i(q=2)-1$ for some $i \in [1,12]$)
have been shown in the Fig. \ref{Figure:5}. Each set corresponds to
a synthetic multifractal rough surface generated according to the algorithm
presented in section II. Our results demonstrate that
the $x_l$ exponent, not only depends on the value of  the Hurst
exponent but also depend on the different sets of $p$ values (see the upper
panel of Fig. \ref{Figure:5} ). In other words, as reported in Table \ref{Tab1} as well as shown in upper panel of Fig. (\ref{Figure:5}),  the sets $i=4$, $i=5$ and $i=6$ of $p$ values have equal Hurst exponent, nevertheless, the corresponding correlation exponents, $x_l$, for these sets differ completely.  On the other hand, at the level of
our numerical accuracy, as shown in the lower panel of Fig.
\ref{Figure:5}, the value for the smoothened multifractal surfaces correlation
exponents is the same as that reported for the mono fractal
rough surfaces, namely $x_l=\frac{1}{2}$.

\begin{figure}[t]
    \centering
        \includegraphics[width=0.50\textwidth]{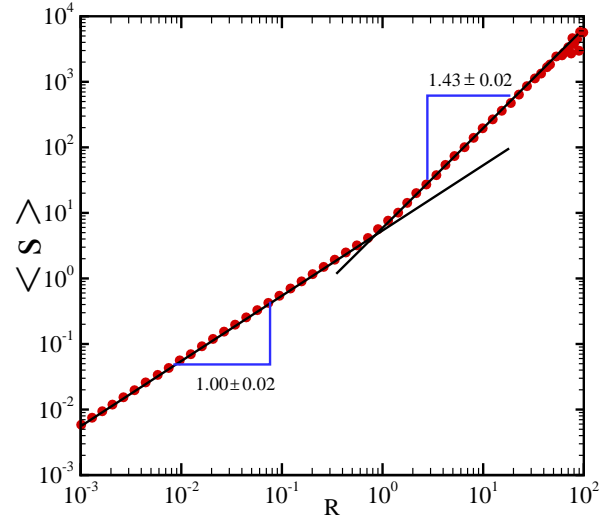}
 \caption{(Color online) The log-log plot of  $\langle s\rangle (R) $ versus $ R $ for synthetic multifractal singular rough surface
for  $H_4=0.608$.} \label{Figure:7}
\end{figure}

\begin{figure}[t]
    \centering
        \includegraphics[width=0.50\textwidth]{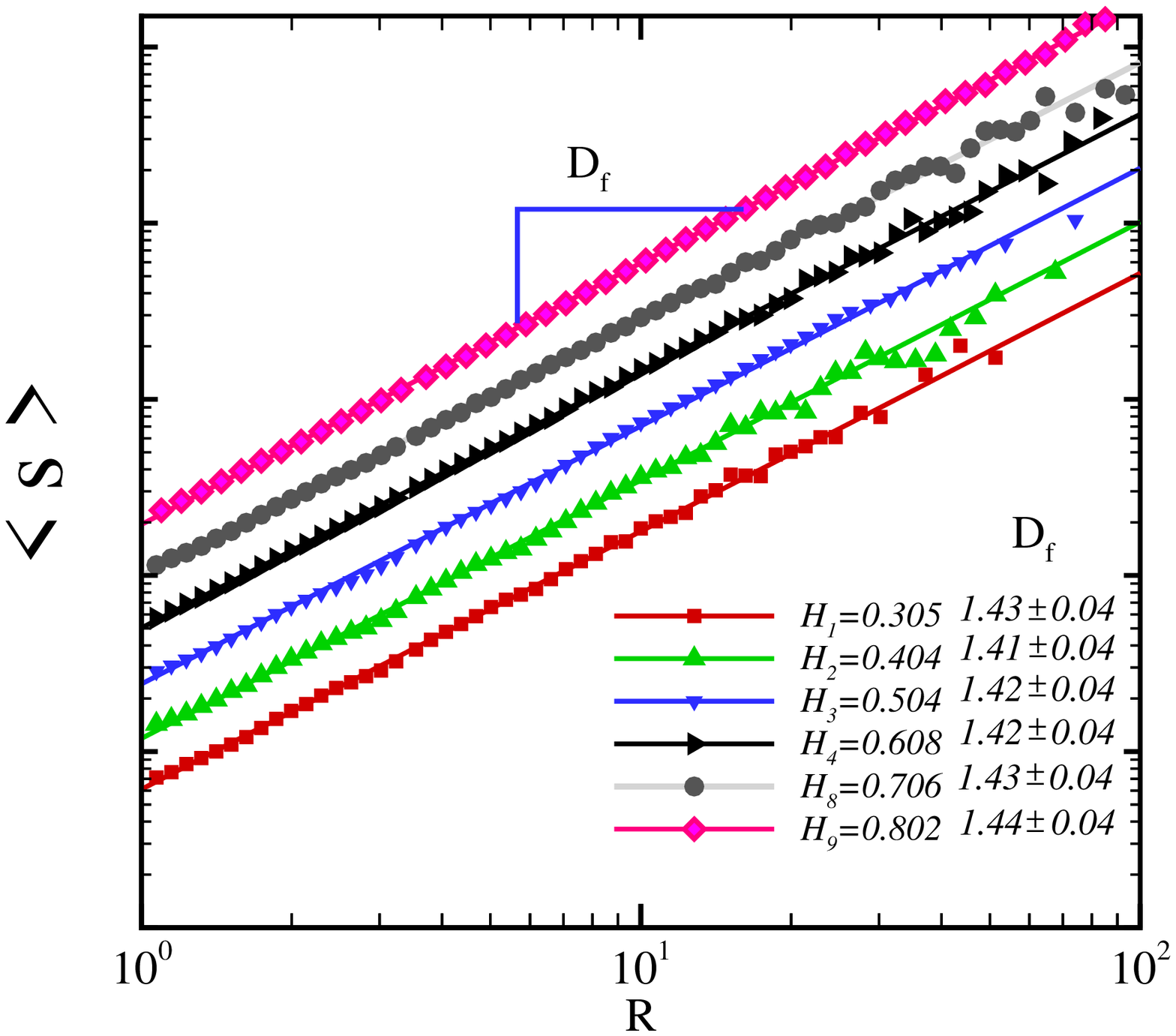}
 \includegraphics[width=0.50\textwidth]{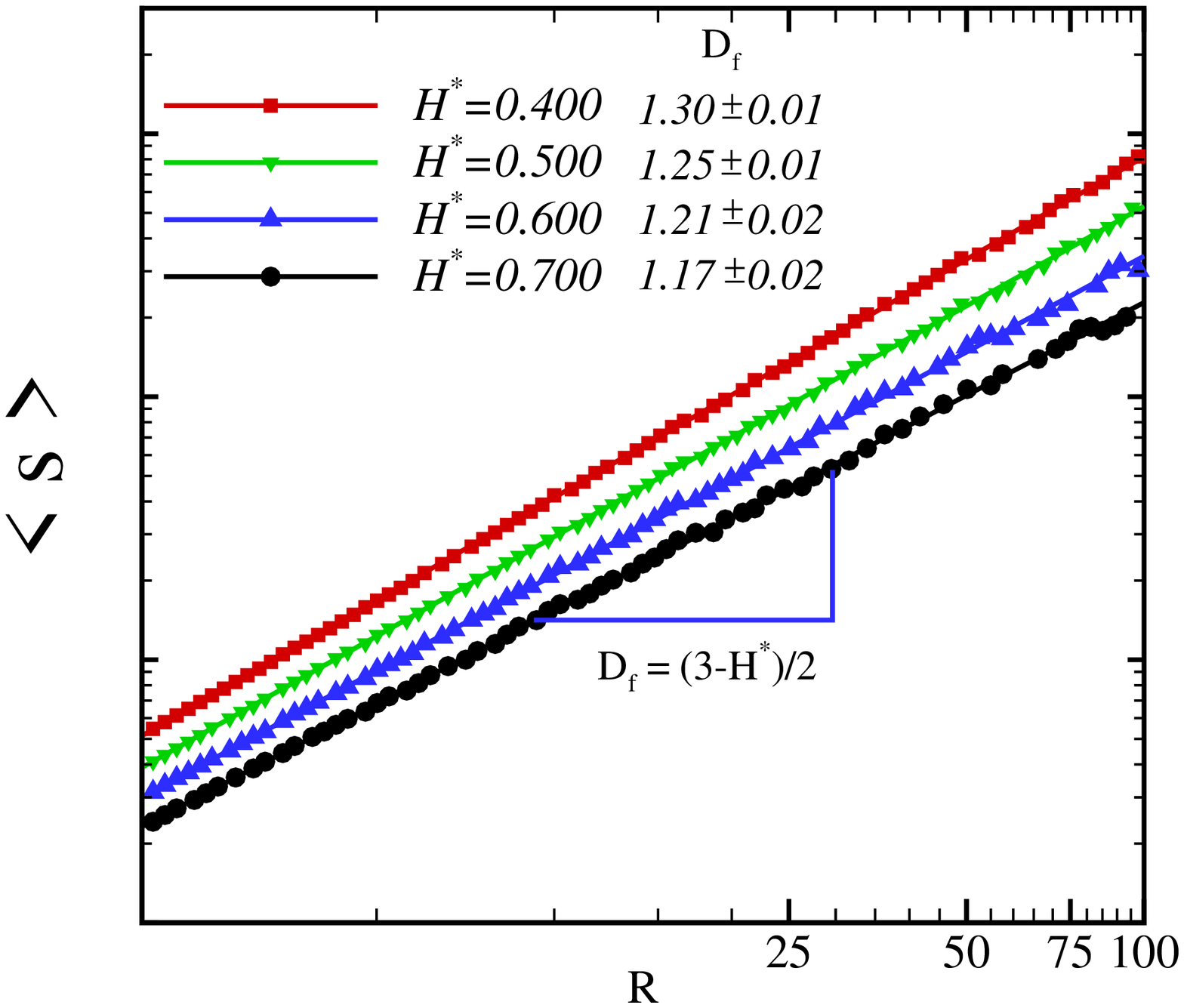}
\caption{(Color online) Upper panel: Log-log of  $\langle
s\rangle(R) $ versus $R $ for singular multifractal rough surfaces for various sets of $p$ values reported in Table I.
Lower panel:  The same diagram  for smoothened synthetic multifractal
rough surfaces.  The sample size is
$4096\times 4096$ and the ensemble average was done over $100$
realizations. To make more sense, we shifted the values of  $\langle
s \rangle$ vertically for different multifractal rough surfaces.}
\label{Figure:8}
\end{figure}

\subsection{Fractal dimension}

To calculate the fractal dimension of a contour loop, we have
calculated the perimeter and radius of gyrations of different
contour loops. Figure \ref{Figure:7} shows log-log plot  of $\langle s\rangle(R)$
versus $\langle R\rangle $ values for synthetic multifractal rough surfaces with typical value of Hurst exponent,
$H_4=0.608$. There are two distinct regions with
different slopes in the diagram; the first region is related to a
large number of small loops with radius smaller than one $(R<1)$
with $D_f=1.00 \pm 0.01$. This is not a relevant phenomenon and it
comes from the contouring algorithm that produces lots of
 contour loops around very small clusters (made
usually from one cell). In the second region $(R>1)$ the slope
increases to $1.43\pm 0.02$ and it keeps to follow the scaling
behavior up to very large sizes. 
The slopes for different Hurst exponent
follow the relation $D_f=(3-H)/2$ for mono-fractal case \cite{KHS}. For various values of
the Hurst exponent our computation is shown in the upper panel
of Fig. \ref{Figure:8} (see also Table \ref{Tab4}). At $1\sigma$
confidence interval all slopes are the same. On the contrary, in the
case of the contour lines of the convolved rough surfaces with
arbitrary  $H^*$'s the fractal dimension of a contour line follows
the formula of a mono fractal surface with $H=H^*$, namely
$D_f=(3-H^*)/2$. It is quite interesting that these results are
completely independent form the $p$ values (lower panel of Fig.
\ref{Figure:8}). This simply means that the fractal dimension of the
contour loops of the singular rough surfaces does not change with
respect to the $h(q=2)$. In other words,  in contrast
to the mono fractal case, $h(q=2)$ alone can not represent the
properties of the underlying singular multifractal rough surface.

\begin{figure}[t]
    \centering
        \includegraphics[width=0.50\textwidth]{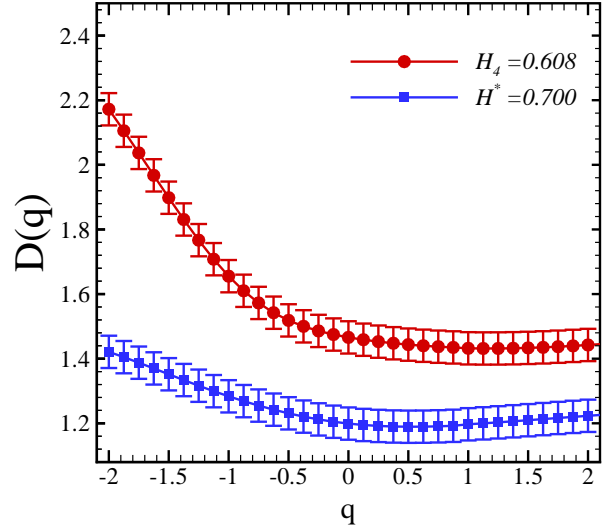}
\caption{(Color online) Generalized fractal dimension versus $q$ for
singular measure with $H_4=0.608$ and that of convolved by
$H^*=0.700$. For singular and smoothened  surfaces $D_f=1.46\pm0.05$ and
$D_f=1.19\pm0.05$, respectively.} \label{Figure:10}
\end{figure}

We also calculated the fractal dimension by using partition function
introduced in Eqs. (\ref{partition11}) and (\ref{generalized fractal
dimension}). Figure \ref{Figure:10} shows $D(q)$ as a function of $q$.
The $q$ dependance of these results confirms that contour loops of
synthetic singular and smooth multifractal rough surfaces are
multifractal. For $q=0$ at $68\%$ confidence interval
$D(q=0)=1.46\pm 0.05$. This is also in agreement with the value
determined by  calculating the scaling behavior of the contour
sizes. In addition as we may expect this diagram demonstrates that
the isoheight contour loops of underlying simulated multifractal
rough surfaces behave as a multifractal feature .

\begin{figure}[t]
    \centering
        \includegraphics[width=0.50\textwidth]{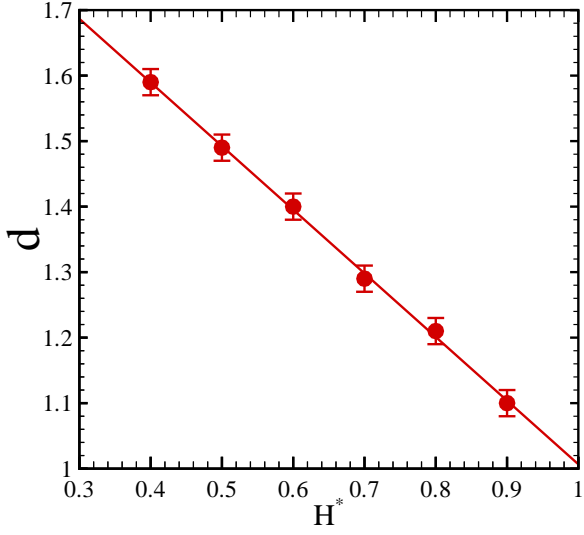}
     \caption{(Color online)  Fractal dimension of all contours of the smoothened multifractal surfaces as a function of $H^*$. The solid line corresponds
     to linear fitting function.} \label{Figure:13}
\end{figure}

As mentioned , the fractal dimension of all the contours, $d$,
differs from the fractal dimension of a contour loop $D_f$. The
fractal dimension of a contour set for mono fractal rough surfaces
is given by $d=2-H$. For the smoothened multifractal rough surfaces
introduced by Eq. (\ref{smooth function}), the fractal dimension of
the contour set is $d=2-H^*$ \cite{wt1}. We have calculated the
fractal dimension of the contour set by using the box counting
method. As previously, we used a least-squares equation (Eq.
(\ref{likelihood1})) to determine the slope in the log-log diagram
of the number of segments that will cover the underlying feature
$N(l)$ versus length scale $l$ for different  Hurst exponents. To
obtain best fit value for the slope corresponding to our  data,  as
well as its error, we divided the data into different ranges and  determined the slope by least-squares method. To do so
according to likelihood function (Eq. (\ref{likelihood1})), we
define $\chi^2$ as
\begin{equation}
\chi ^2(d)=\sum_{i=1} ^{\mathcal{N}}
\frac{[N(l_i)-N_{\rm the.}(l_i;d)]^2}{\sigma (l_i)^2},
\end{equation}
where ${\mathcal{N}}$ is the number of partitioning, namely
${\mathcal{N}}=L/l_{\mathcal{N}}$,  $N_{\rm the.}(l_i;d)\sim
l_{i}^{-d}$ and $\sigma^2 (l_{i})$ is the variance of the data in
the corresponding range. Finally we determined the minimum $\chi ^2$
and  the best slope for the data. Figure \ref{Figure:13} corresponds
to synthetic smoothened multifractal rough surfaces. In addition we
checked that whether the result associated to the smoothened rough
surfaces depends on the set of $p$ values correspond to the  same
value of $h(q=2)$. Our findings confirm that $d$ doesn't depend on
different sets of $p$ values. However, for the singular measure, $d$
depends on the value of $H$ and even $p$'s used for the cascade
algorithm. It  has no regular behavior with respect to $h(q=2)$.
Moreover for various sets of $p$ values giving the same value of
$h(q=2)$,  one finds out different values for fractal dimension of all
contour sets. This is quite surprising because for singular measure
multifractal surface,  we have $H^*=0$ and, therefore,  if the formula
$2-H^*$ was correct in this regime,  we should have $d=2$ for all
the different $h(q=2)$'s. We are not aware of any theoretical
argument that can explain this phenomenon.

\begin{figure}[t]
    \centering
        \includegraphics[width=0.50\textwidth]{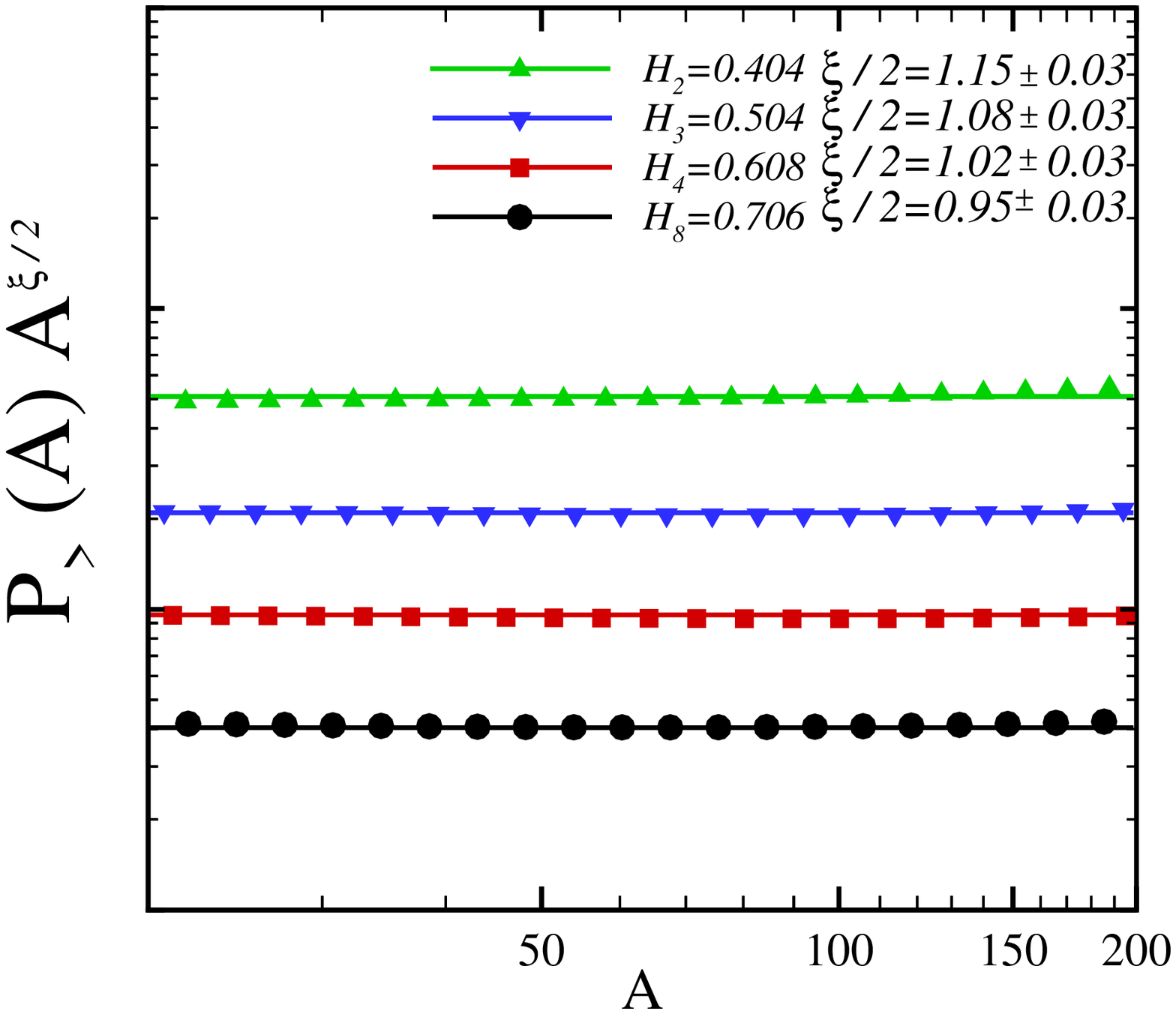}
        \includegraphics[width=0.50\textwidth]{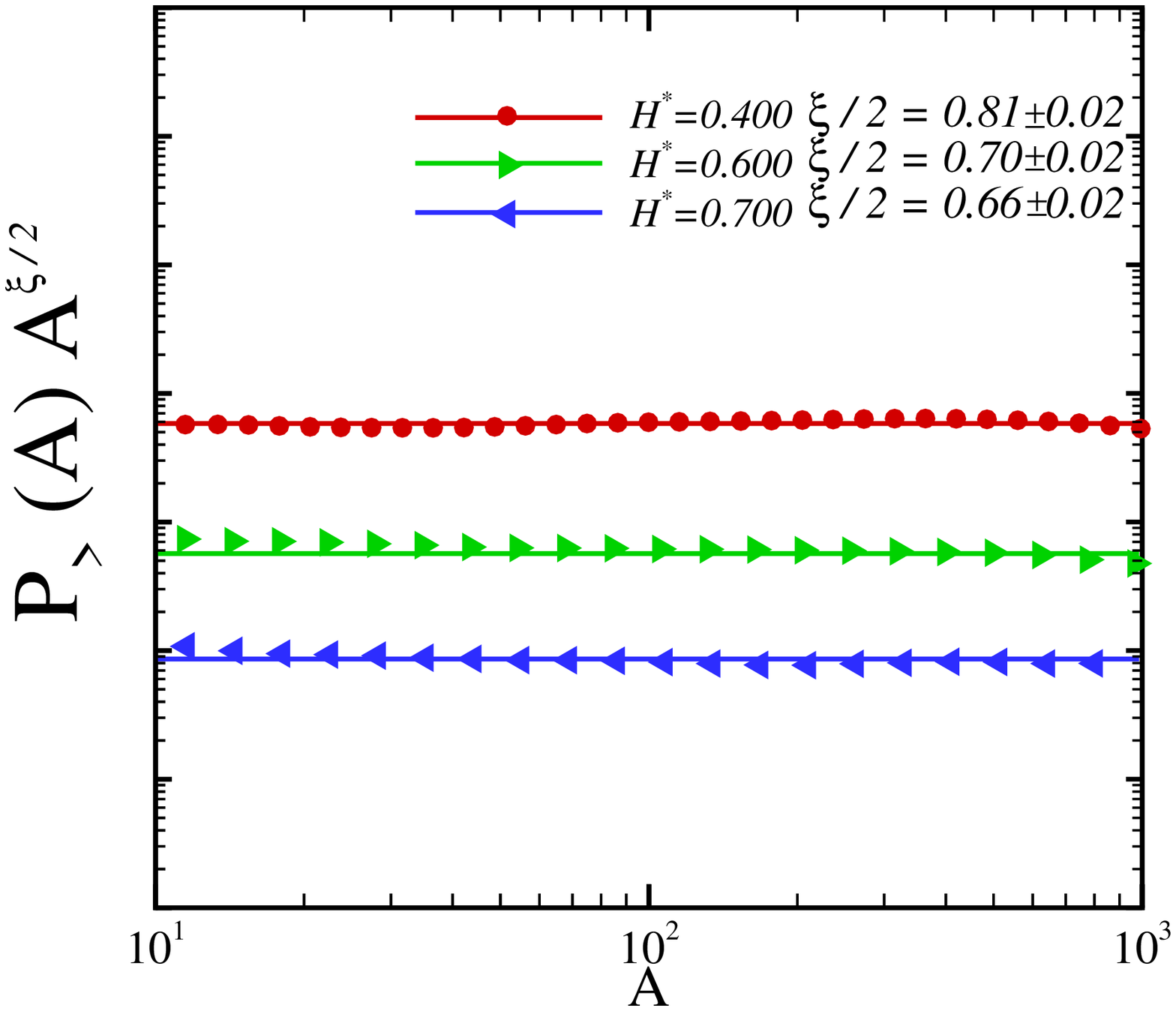}
\caption{(Color online)  Upper panel: The cumulative distributions
of the areas of the contour loops with respect to the area for the
singular  multifractal rough surfaces. The corresponding set of $p$ values is given in Table I.  Lower panel: The same
distribution for the smoothened multifractal surfaces. For clarity, we shifted the value of $y$ axis vertically for both
diagrams.} \label{Figure:11}
\end{figure}
\begin{figure}[t]
    \centering
        \includegraphics[width=0.50\textwidth]{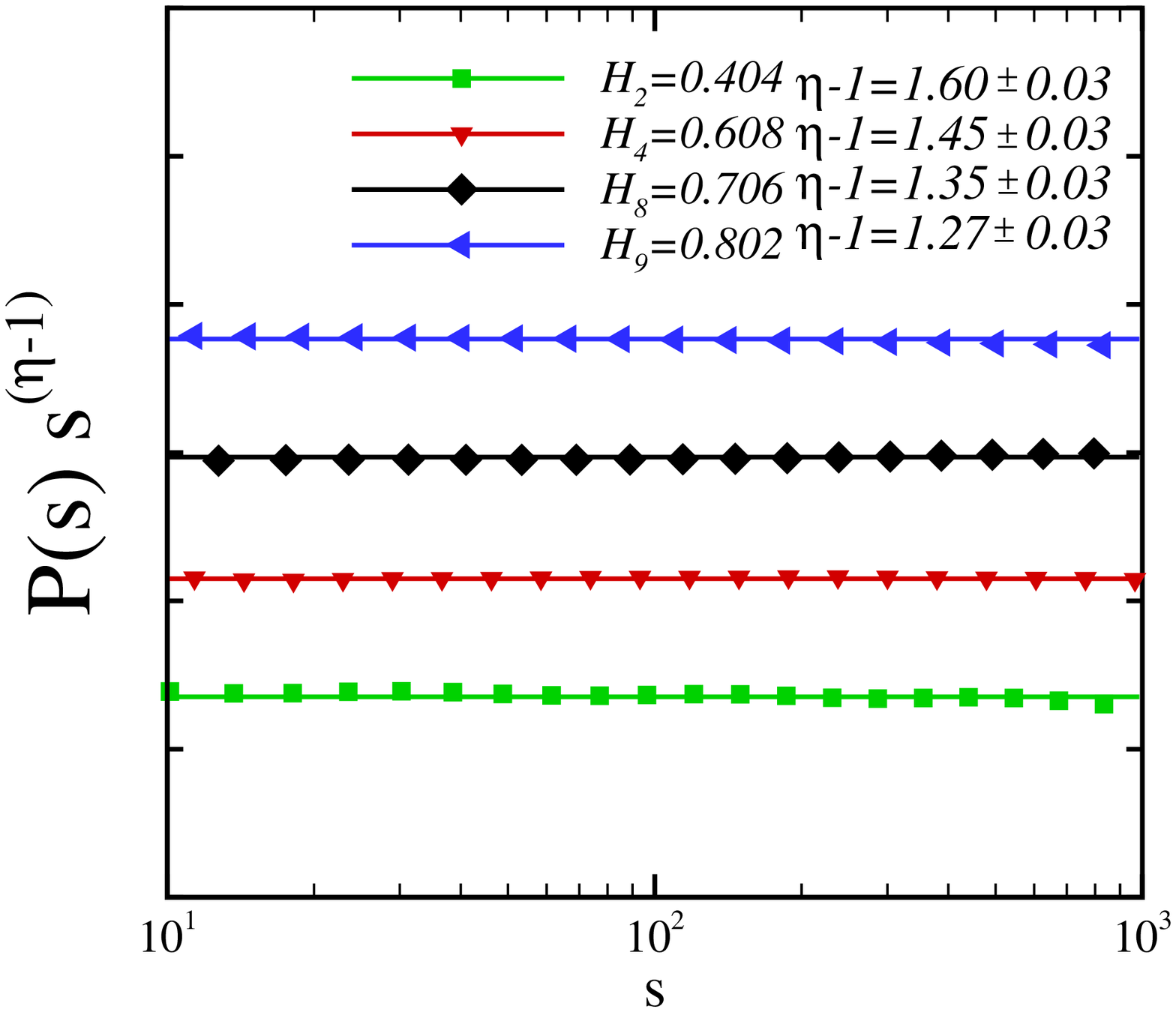}
        \includegraphics[width=0.50\textwidth]{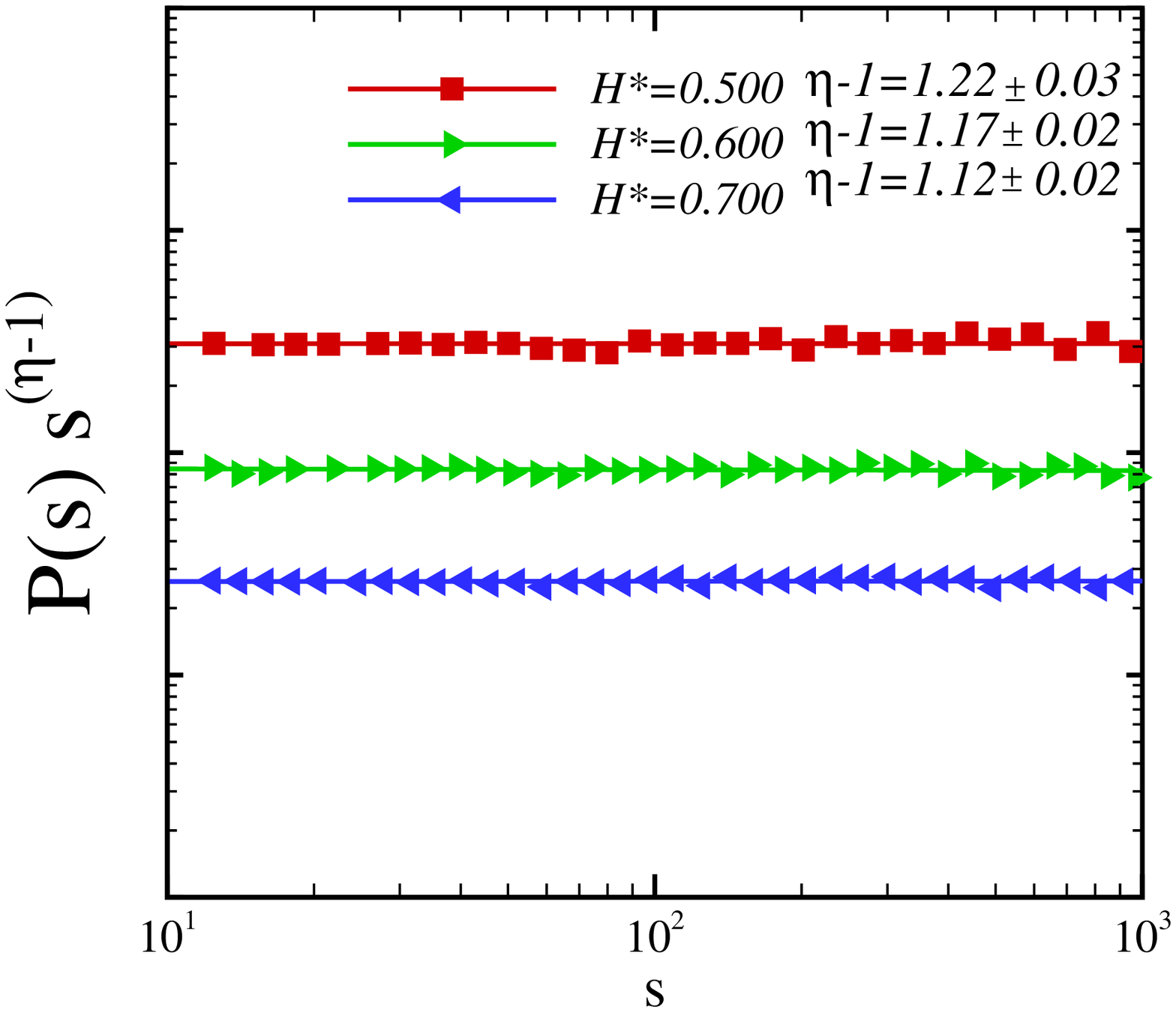}
\caption{ (Color online) Upper panel:The perimeter distribution exponent for
different sets of $p$ values of the singular measure. Lower panel the
same measure for the smoothened multifractal rough surfaces. The values
of $y$ axis are shifted vertically.} \label{Figure:12}
\end{figure}

\subsection{Cumulative distribution of areas}

To calculate the exponent $\xi$ we have calculated the
$P_{>}(A)A^{\xi/2}$ with respect to the area of the contour loops.
In Fig. \ref{Figure:11} and table \ref{Tab4} we have shown the
results for various values of Hurst exponent reported in table
\ref{Tab1} and averaging is done over 100 realizations.  The results
are quite different from what we expect for mono-fractal rough
surfaces. For mono fractal rough surfaces we have $\xi=2-H$. It must
be pointed out that for synthetic singular multifractal rough
surfaces $\xi$ decreases by increasing $H$,  which is the same as
mono-fractal rough surfaces.  In addition  $\xi$ not only depends on
$h(q=2)$ but also is affected by other values of $h(q)$'s. This finding
is due to the multifractality nature of the singular measure rough
surface. The same computation for the smoothened multifractal rough
surfaces is shown  in lower panel of Fig. \ref{Figure:11}.
This results confirm that the exponent  is controlled by $H^*$, and
$\xi$ is given by the same equation as for the mono fractal rough
surfaces.

\subsection{Probability distribution of contour length}

Final remark concerns the probability distribution of contour
length.  To this end we investigated the logarithmic diagram of
$P(s)s^{\eta-1}$ versus $s$. We have depicted the results for the
synthetic singular as well as smoothened multifractal surfaces for
various values of $h(q=2)$. For the smoothened multifractal rough
surfaces again the exponents follow the behavior of the mono fractal
surfaces (see Fig. 14).

In spite of  the huge difference between the geometrical exponents of the contour loops of mono-fractal
rough surfaces and singular multifractal rough surfaces,
the hyperscaling relations $\frac{\xi}{D_f}=\eta-1$ and $D_f=\frac{2x_l-2}{\eta-3}$ are valid up to numerical
accuracy (see  Table. \ref{Tab4} and Table. \ref{Tab5}). 
The important factors in obtaining this hyperscaling relation concern power-law relations for $\tilde{P}(s)$ and $P_{>}(A)$.
The second hyperscaling relation comes from the following equality
\begin{eqnarray}\label{hyperproof2}
\int_{0}^{R}G(r)d^2r\sim\int_{0}^{\infty}\min(s,R^{D_f})P(s)ds.
\end{eqnarray}
Both sides are proportional to the mean of the length of that portion of the contours passing through origin
which lies within a radius $R$ from the origin \cite{KH}. 

\begin{table}[t]

\begin{center}
\begin{tabular}{|c|c|c|c|c|c|}\hline
 $H$   & $\eta$ & $D_{f}$ & $\xi$&$2x_l$ \\
\hline
$H_1=0.305$ & $2.67\pm0.03$ & $1.43\pm0.04$ &$2.44\pm0.06$&$1.60\pm0.10$\\
\hline
$H_2=0.404$ & $2.60\pm0.03$ & $1.41\pm0.04$ &$2.30\pm0.06$&$1.49\pm 0.10$\\
\hline
$H_3=0.504$ & $2.50\pm0.03$ & $1.42\pm0.04$ &$2.16\pm0.06$&$1.25\pm0.05$\\
\hline
$H_4=0.608$ & $2.45\pm0.02$ & $1.42\pm0.04$ &$2.04\pm0.06$&$1.30\pm0.03$\\
\hline

$H_5=0.608$ & $2.74\pm0.02$ & $1.43\pm0.04$ &$2.50\pm0.06$&$1.70\pm0.03$\\

\hline
$H_6=0.608$ & $2.64\pm0.02$ & $1.42\pm0.04$ &$2.31\pm0.06$&$1.53\pm0.03$\\


\hline
$H_8=0.706$ & $2.35\pm0.02$ & $1.43\pm0.04$ &$1.90\pm0.06$&$1.12\pm 0.03$\\
\hline
$H_9=0.802$ & $2.27\pm0.02$ & $1.44\pm0.04$ &$1.80\pm0.06$&$1.02\pm 0.03$\\
\hline
    \end{tabular}
\end{center}
\caption{\label{Tab4}Different geometrical exponents of the
contour loops extracted from surfaces with  different sets of $p-$values reported in Table I. Theses values completely dependent on the $p$ values.}
\end{table}

\begin{table}

\begin{center}
\begin{tabular}{|c|c|c|c|c|c|}\hline
 $H$   & $\eta-1$ & $\frac{\xi}{D_{f}}$ & $3D_f+2x_l$&$D_f\eta+2$ \\
\hline
$H_1=0.305$ & $1.67\pm0.03$ & $1.71\pm0.06$ &$5.89\pm0.16$&$5.82\pm 0.12$\\
\hline
$H_2=0.404$ & $1.60\pm0.03$ & $1.63\pm0.06$ &$5.72\pm0.16$&$5.67\pm0.12$\\
\hline
$H_3=0.504$ & $1.50\pm0.03$ & $1.52\pm0.06$ &$5.51\pm0.13$&$5.55\pm0.11$\\
\hline
$H_4=0.608$ & $1.45\pm0.02$ & $1.44\pm0.06$ &$5.56\pm0.12$&$5.48\pm0.10$\\
\hline

$H_5=0.608$ & $1.74\pm0.02$ & $1.75\pm0.06$ &$5.99\pm0.12$&$5.92\pm0.11$\\
\hline

$H_6=0.608$ & $1.64\pm0.02$ & $1.63\pm0.06$ &$5.79\pm0.12$&$5.75\pm0.11$\\
\hline


$H_8=0.706$ & $1.35\pm0.02$ & $1.33\pm 0.06$ &$5.41\pm0.12$&$5.36\pm0.10$\\
\hline
$H_9=0.802$ & $1.27\pm0.02$ & $1.25\pm0.05$ &$5.34\pm0.12$&$5.27\pm0.10$\\
\hline
    \end{tabular}
\end{center}
\caption{\label{Tab5}Verification of two basic hyperscaling relations for synthetic singular measure multifractal
rough surfaces.}
\end{table}

\begin{table}

\begin{center}

\begin{tabular}{|c|c|c|}
\hline  Exponent & Singular measure & Smoothened multifractal   \\
\hline  $x_l$   &{\rm Depends on $p-$values}  & $\frac{1}{2}$ \\
\hline  $D_f$&{\rm Identical}  & $\frac{3-H^*}{2}$ \\
\hline  $D(q)$&{\rm Depends on $q$}  & {\rm Depends on $q$} \\
\hline  $d$&{\rm Depends on $p-$values}   &  $2-H^*$\\
\hline  $\eta-1$& {\rm Depends on $p-$values}  & $\frac{4-2H^*}{3-H^*}$  \\
\hline  $\xi$&  {\rm Depends on $p-$values} &  $2-H^*$\\
\hline $\frac{\xi}{D_f}=\eta-1$&{\rm YES} &{\rm YES} \\
\hline $D_f=\frac{2x_l-2}{\eta-3}$&{\rm YES}&{\rm YES}\\
\hline
   \end{tabular}
\end{center}
\caption{\label{res} A summarization of results given in this paper based on contouring analysis for synthetic singular and smoothened multifractal rough surfaces.}
\end{table}

\section{Conclusion}

In this paper we have studied  the contour lines of particular
multifractal rough surfaces, namely the so-called multiplicative
hierarchical cascade $p$ model. Utilizing a stochastic cascade method
\cite{Feder,Biolog}, singular measure (original) and smoothened
(convolved) multifractal rough surfaces with different Hurst
exponents were generated. The $h(q)$ spectrums of these
two dimensional surfaces were determined by MF-DFA method
\cite{Biolog}. Then use of two different algorithms we generated the
contour loops of the systems. Many different geometrical exponents,
such as the fractal dimension of a contour loop, $D_f$ the fractal
dimension of the contour set, $d$, the cumulative distributions of
perimeters and areas and the correlation exponents, $x_l$, were
calculated for the singular and smoothened multifractal surfaces by
use of different methods.
\\
We summarize the most important results given in this
study as follows. 
Our results confirmed that, the exponent of loop correlation
function, $x_l$, for multifractal singular measure, depends on
$p$ values. On the contrary,  for multifractal smoothened surfaces, this
value behaves the same as that of given for mono fractal
rough surfaces (see Fig. \ref{Figure:5}). \\
The scaling exponent of the size of the contours as a function of the radius
representing fractal dimension, $D_f$, is similar for various singular multifractal
rough surfaces. But the relation between $D_f$ and $H^*$ for convolved multifractal
surfaces is similar to mono fractal surfaces. Nevertheless the contour loops have
multifractal nature (see Figs. \ref{Figure:8} and \ref{Figure:10}).\\
The exponent of cumulative distribution of areas, $\xi$, for
singular measure has multifractal nature. But for smoothened surface,
this quantity is controlled by $H^*$ according to $\xi=2-H^*$ and is
completely independent of $p$ values
(see Fig. \ref{Figure:11}). \\
Consequently, in the case of singular measure surfaces, all of the
exponents show significant deviations from the well-known formulas
for the mono-fractal rough surfaces. They depend on the generalized
Hurst exponents $h(q)$, whreas for convolved multifractal surfaces,
all geometrical exponents are controlled by $H^*$ according to a
mono fractal system. We emphasize  that interestingly, the
hyperscaling relations, namely, $\frac{\xi}{D_f}=\eta-1$ and
$D_f=\frac{2x_l-2}{\eta-3}$ at $1\sigma$ confidence interval, are
valid for both singular and smoothened multifractal
rough surfaces
(see Table. \ref{Tab4}, \ref{Tab5}).
In this system which is labeled by $H^*$, many relevant properties
are controlled by a few relations that have been presented for
mono-fractal cases. However singular and smoothened multifractal
surfaces have multifractal nature but using geometrical analysis,
they belong to different class which is a non-trivial result. Table \ref{res} contains most important results given in this paper.

Finally, to make more complete this study, it is useful to extend this approach for various
simulated rough surfaces by use of different methods
and examine their hyperscaling relations. In addition, there are some methods
to distinguish various  multiplicative cascade methods such as $n$-point statistics \cite{martin98}.

\textbf{Acknowledgments:}

We are grateful to M. Ghaseminezhad for cross checking some of the results and also for fruitful discussions.
MSM and SMVA are grateful to associate office of ICTP and their hospitality. The work of SMVA was supported
in part by the Research Council of the University of Tehran.

\end{document}